\documentclass[fleqn,usenatbib]{mnras}

\usepackage{newtxtext,newtxmath}

\usepackage[T1]{fontenc}

\DeclareRobustCommand{\VAN}[3]{#2}
\let\VANthebibliography\thebibliography
\def\thebibliography{\DeclareRobustCommand{\VAN}[3]{##3}\VANthebibliography}

\usepackage{graphicx}	
\usepackage{fixltx2e}
\MakeRobust{\overrightarrow}

\title[Coronal temperature variation in Seyfert galaxies]{Search for coronal temperature variation in Seyfert galaxies}

\author[Pal et al.]{
Indrani Pal$^{1,2}$\thanks{E-mail: indrani.pal.iiap.res.in}
and C. S. Stalin $^{1}$
\\
$^{1}$Indian Institute of Astrophysics, Block II, Koramangala, Bangalore 560 034\\
$^{2}$Pondicherry University, R.V. Nagar, Kalapet, 605014, Puducherry, India.\\
}

\date{Accepted XXX. Received YYY; in original form ZZZ}

\pubyear{2022}

\begin{document}
\label{firstpage}
\pagerange{\pageref{firstpage}--\pageref{lastpage}}
\maketitle

\begin{abstract}
While the temperature of the X-ray corona ($\rm{kT_e}$) in active galactic nuclei (AGN) are known for many sources, its variation if any is limited to a handful of objects. This is in part due to the requirement of good signal-to-noise X-ray spectra covering a wide range of energies. We present here results on the X-ray spectral analysis of 18 Seyferts, having more than one epoch of observations to look for variation in $\rm{kT_e}$. The data for a total of 52 epochs on these 18 AGN  were taken from observations carried out by {\it NuSTAR} in the 3$-$79 keV energy band. From phenomenological and physical model fits to the multi epoch data on these 18 sources from {\it NuSTAR}, we could constrain the cut-off energy ($E_{cut}$) in a large fraction of the sources. Also, from Comptonized model fits we could obtain $\rm{kT_e}$ for our sample. Of the 18 sources, at the 90\% confidence level, evidence for variation in $\rm{kT_e}$ was found for only one source namely MCG+08-11-011. For this source between two epochs, separated by about five years, we found $\rm{kT_e}$ to decrease from  57$^{+29}_{-16}$ keV to 30$^{+11}_{-7}$ keV. During the same period, the flux decreased from (12.60 to 14.02) $\times$ 10$^{-11}$ erg cm$^{-2}$ s$^{-1}$ and the optical depth increased from 1.68 to 2.73. We thus found a positive correlation between flux and coronal temperature with a reduction of about 40\% in optical depth. Our observations tend to favour the vertically outflowing corona scenario for the observed variation in  $\rm{kT_e}$ in MCG+08-11-011.
\end{abstract}

\begin{keywords}
galaxies: active -- galaxies: nuclei -- galaxies: Seyfert -- X-rays: galaxies
\end{keywords}



\section{Introduction}
\label{sec:introduction}
Active galactic nuclei (AGN) are the highly luminous ($\sim 10^{42} - 10^{48}$ erg s$^{-1}$)  central region of galaxies. 
They are believed to be powered by accretion of matter onto supermassive black 
holes (SMBHs) with masses of $\sim$ 10$^6$ $-$ 10$^{10}$ $M_{\odot}$ at the centres of 
galaxies \citep{1969Natur.223..690L,1973A&A....24..337S,1984ARA&A..22..471R,1993ARA&A..31..473A,1995PASP..107..803U}. They emit
radiation over the entire accessible electromagnetic spectrum, which is also
found to be variable \citep{1995ARA&A..33..163W}. In particular, X-ray emission is ubiquitously present in various 
classes of AGN \citep{1978MNRAS.183..129E,1993ARA&A..31..717M}, though the dominant physical processes that produce the observed
X-ray emission may be different among the different categories of AGN.

The variable hard X-ray ($>$2 keV) emission observed commonly in the radio-quiet 
category of AGN is believed to originate in a hot and 
compact region called the corona \citep{1991ApJ...380L..51H, 1994ApJ...432L..95H} 
situated close to the vicinity of the SMBH. Observations point to the corona to be
compact located within 3$-$10 gravitational radii ($\rm{R_{g}}$) of the SMBH 
\citep{2015MNRAS.451.4375F}. 
The observed X-ray emission in 
AGN is thus due to thermal Comptonization of the accretion disk photons 
by the hot electrons ($\rm{T_e \sim 10^{8-9}}$ K) in the corona leading to a power law 
shaped X-ray spectrum with a high energy cut-off ($\rm{E_{cut}}$;\citealt{1980A&A....86..121S}). The observed
X-ray spectral power-law slope depends on the optical depth ($\tau$), temperature of the coronal 
plasma ($\rm{kT_{e}}$) and the seed photon temperature. Thus, by analyzing the 
hard X-ray spectra of AGN, one can constrain $\tau$ and ${\rm kT_e}$. The primary X-ray 
continuum emission from the corona also gets reprocessed via Compton reflection 
from the accretion disk or more distant material giving rise to the reflection 
hump at around 15$-$30 keV and also the broad FeK$\alpha$ line at 
6.4 keV \citep{1991MNRAS.249..352G, 1993MNRAS.262..179M}. The broad
FeK$\alpha$ line is also known to show relativistic effects as it originates
close to the central SMBH. The rapid spin of the SMBH in AGN can lead to
strong red wing of the FeK$\alpha$ line that can 
extend down to 3-4 keV \citep{2021iSci...24j2557C}, which has also been used to 
measure the spin of SMBH in many AGN \citep{2019NatAs...3...41R}. In addition to 
the broad FeK$\alpha$ line, 
narrow FeK$\alpha$ line component is also seen in some AGN \citep{2010ApJS..187..581S}  
whose origin is likely to be from
the material at larger distance from the SMBH.  

In spite of the 
corona being an integral part of AGN, details such as its nature, geometry, 
location and the physical mechanisms that power it are still unknown.  
The analysis of the different spectral features present in the hard X-ray spectra 
of AGN will help constraining the geometry of the X-ray emission region and 
explaining the physical processes that lead to the origin of these components. 
For disentangling the properties of each of the spectral components and constraining 
their characteristic parameters, there is a requirement of thorough analysis of 
high signal-to-noise (S/N) ratio AGN spectra. In particular, measurements of the
photon index of the underlying power law continuum and the $E_{cut}$ 
of the continuum can be of help in constraining the characteristics of the 
X-ray corona.

The launch of the {\it Nuclear Spectroscopic Telescope Array} ({\it NuSTAR}; 
\citealt{2013ApJ...770..103H}) in the year 2012 has started to provide the 
observational leads (through the measurements of $\rm{E_{cut}}$) to our understanding 
of the corona in AGN, owing to its high sensitivity beyond 10 keV. Importantly, 
in addition to the determination of $\rm{E_{cut}}$, there are also evidences for 
the variation in the $\rm{E_{cut}}$ values from the studies of the multi-epoch 
{\it NuSTAR} data. For example, in eight sources, namely MCG-5-23-16 
\citep{2017ApJ...836....2Z}, 3C 382 \citep{2014ApJ...794...62B}, NGC 4593 
\citep{2016MNRAS.463..382U}, NGC 5548 \citep{2015A&A...577A..38U}, Mrk 335 \citep{2016MNRAS.456.2722K}, 4C +74.26 \citep{2018ApJ...863...71Z}, NGC 3227 \citep{2021MNRAS.tmp...68K} and SWIFT J2127.4+5654 \citep{2021MNRAS.tmp...68K} variations in the $\rm{E_{cut}}$ values are available in the literature.  According to 
Comptonization models, for a hot corona with slab geometry, $\rm{E_{cut}}$ = 2$-$3 $\rm {kT_e}$ 
for optically thin and thick plasma respectively \citep{2001ApJ...556..716P}.

In all the above cases where variation in $\rm{E_{cut}}$ are reported in the literature, for a majority of the sources, ${\rm E_{cut}}$ were  measured from phenomenological 
model fits to the observed X-ray spectra.  However,  owing to a variety of complex 
reasons behind the variation in $\rm {kT_e}$, fitting the data with a phenomenological 
cut-off power law does not allow for a unique interpretation \citep{2015A&A...577A..38U}. 
Also, from a recent study, \cite{2019A&A...630A.131M} have shown that the commonly 
adopted relation of $\rm{E_{cut}}$ = 2$-$3 $\rm{kT_{e}}$ \citep{2000ApJ...540..131P} 
is not valid for all values of $\tau$ and $\rm{kT_{e}}$, instead valid for only 
low values of $\tau$ and $\rm{kT_{e}}$. In the case of a non-static corona such as the 
one with outflows, this conventional relation between $\rm{E_{cut}}$ and $\rm{kT_{e}}$ 
can be complicated \citep{2014ApJ...783..106L}. Recently, using the multi epoch 
data from the {\it NuSTAR} archives for three sources
namely NGC 3227, MR 2251$-$178 and NGC 5548, , \cite{2022A&A...662A..78P} found variation in $\rm{kT_{e}}$ in the source NGC 3227. 
From physical model fits to the multi epoch data, \cite{2020MNRAS.492.3041B} 
confirmed the change in $\rm{kT_{e}}$ in another Seyfert galaxy, namely, 
Ark 564. Recently \cite{2021MNRAS.tmp...68K} also reported the variation of 
$\rm{kT_{e}}$ in NGC 3227 and SWIFT J2127.4+5654. Though in ESO 103-035, 
\cite{2021ApJ...921...46B} found $\rm{kT_e}$ to increase with the brightness of 
the source, in Ark 564, \cite{2020MNRAS.492.3041B} found 
$\rm{kT_e}$ to decrease with source brightness. Thus the limited results
available in the literature, seem to indicate that the variation of $\rm{kT_e}$ with 
source brightness may be different among sources. To confirm or refute these findings known on 
limited sources, a systematic and homogeneous analysis of the variation of $\rm{kT_e}$ and
its correlation with various physical properties need to be carried out on a large number
of sources. A limitation to the study of such kind, is the availability of  multiple epochs of 
observations on a source. We have carried out an investigation of the variation in $\rm{kT_e}$  in
18 AGN. The choice of sources is driven by the availability of multiple epochs of
data from {\it NuSTAR}. The selection of our sample and data reduction are described in 
Section 2, model fits to the data are described in Section 3 and the results and discussion are given in Section 4, followed by the Summary in the final section.

\begin{table*}
\caption{Details of the sources analyzed in this work.  The columns are (1) name of the source, (2) right ascension ($\alpha_{2000}$) (h:m:s), (3) declination ($\delta_{2000}$) (d:m:s), (4) redshift ($z$), (5) galactic hydrogen column density ($\rm N_{H}^{gal}$) in units of $10^{22}$ atoms $\rm{cm^{-2}}$, (6) type of the source, (7) observation ID, (8) epoch, (9) date of observation, and (10) exposure time in sec. The information such as the right ascension, declination, $z$ and type of the source are from NED (\href{https://ned.ipac.caltech.edu/}{https://ned.ipac.caltech.edu/}).} \label{Table-1}
\centering
\begin{tabular}{p{0.12\linewidth}p{0.09\linewidth}p{0.11\linewidth}p{0.08\linewidth}p{0.06\linewidth}p{0.03\linewidth}p{0.10\linewidth}p{0.03\linewidth}p{0.10\linewidth}p{0.07\linewidth}}
\hline
Source & $\alpha_{2000}$ & $\delta_{2000}$ & $z$ & $\rm N_{H}^{gal}$ & Type & OBSID & Epoch & Date & Exposure \\
\hline
1H0419$-$577         & 04:26:00.72 & $-$57:12:00.97   & 0.104  & 0.013  & Sy1.5 & 60402006002 & A & 2018-05-15 & 64216  \\
                 &  &            &             &                &       &  60402006004 & B & 2018-11-13 & 48273  \\ 
                 &  &           &             &                &       & 60101039002 & C & 2015-06-03 & 169462    \\
Mrk 915         & 22:36:46.50 &  $-$12:32:42.58   & 0.025  & 0.064  & Sy1.8 & 	  60002060002 & A & 2014-12-02  & 52977 \\
                 & & & &  &                                                &  60002060004  & B & 2014-12-07 & 54249    \\
                & & & &   &                                                &  60002060006 & C & 2014-12-12  & 50678 \\
3C 111  &  04:18:21.27  &  +38:01:35.80 &  0.049 &  0.434 & Sy1 & 	60202061002 & A & 2017-12-26 & 21324  \\
        & & & & & &                                                 60202061004 & B & 2017-12-29 & 49361  \\
        & & & & & &                                                 60202061006 & C & 2019-01-04 & 51697  \\
NGC 3783 & 11:39:01.76 & $-$37:44:19.21 & 0.097 & 0.138 & Sy1.5 &  60101110002 & A & 2016-08-22 & 41265 \\
& & & & & &                                                      60101110004 & B & 2016-08-24 & 42428 \\
& & & & & &                                                      80202006002 & C & 2016-12-11 & 25657 \\
& & & & & &                                                      80202006004 & D & 2016-12-21 & 25367 \\
NGC 7469 & 23:03:15.62 & +08:52:26.39 & 0.0163 &  0.052 & Sy1.2 & 60101001002 & A & 2015-06-12 & 21579 \\
& & & & & &                                                         60101001004 & B & 2015-11-24 & 20027 \\
& & & & & &                                                         60101001006 & C & 2015-12-15 & 22521 \\
& & & & & &                                                         60101001008 & D & 2015-12-22 & 23484 \\
& & & & & &                                                         60101001010 & E & 2015-12-25 & 20847 \\
& & & & & &                                                         60101001012 & F & 2015-12-27 & 20963 \\
& & & & & &                                                         60101001014 & G & 2015-12-28 & 23398 \\
Mrk 110 & 09:25:12.87 & +52:17:10.52 & 0.035 & 0.014 & Sy1 & 60201025002 & A & 2017-01-23 & 184563 \\
& & & & & &                                                  60502022002 & B & 2019-11-16 & 86772 \\
& & & & & &                                                  60502022004 & C & 2020-04-05 & 88647 \\
UGC 06728 & 11:45:16.02 & +79:40:53.42 & 0.007 & 0.055 & Sy1.2 & 60160450002 & A & 	2016-07-10 & 22615 \\
& & & & & &                                                      60376007002 & B &  2017-10-13 & 58077 \\
NGC 4258 & 12:18:57.50 & +47:18:14.30 & 0.002 & 0.016 & Sy1.9  & 60101046002 & A & 2015-11-16 & 54783 \\
& & & & & &                                                      60101046004 & B & 2016-01-10 & 103612 \\
KUG 1141+371 & 11:44:29.87 & +36:53:08.64 & 0.038 & 0.018 &  Sy1 & 60160449002 & A & 2019-12-26 & 21525 \\
& & & & & &                                                       90601618002 & B & 2020-05-13 & 38562 \\
MCG-06-30-15 & 13:35:53.70 & $-$34:17:43.94 & 0.008 & 0.047 & Sy1.2 & 60001047002 & A & 2013-01-29 & 23267 \\
& & & & & &                                                          60001047003 & B & 2013-01-30 & 127219 \\
& & & & & &                                                          60001047005 & C & 2013-02-02 & 29643 \\
NGC 5506 & 14:13:14.89 & $-$03:12:27.28 & 0.006 & 0.048 & Sy1.9 & 	60061323002 & A & 	2014-04-01  & 56585 \\
& & & & & &                                                         60501015002 & B & 2019-12-28 & 	61384 \\
& & & & & &                                                         60501015004 & C & 	2020-02-09 & 47480 \\
MCG+08-11-011 & 05:54:53.61 & +46:26:21.61 & 0.020 & 0.293 & Sy1.5 & 60201027002 & A & 2016-08-16 & 97921 \\
& & & & & &                                                          90701640002 & B & 	2021-12-18 & 25000 \\
GRS 1734-292 & 17:37:28.35 & $-$29:08:02.50 & 0.021 & 0.876 & Sy1 & 60061279002 & A & 2014-09-16 & 20288 \\
& & & & & &                                                        60301010002 & B & 2018-05-28 & 26020 \\
Mrk 926 & 23:04:43.48 & $-$08:41:08.62 & 0.047 & 0.125 & Sy1.5 & 60201029002 & A & 2016-11-21 & 106201 \\
& & & & & &  60761009002 & B &  2021-07-04 & 16688 \\
Mrk 841 & 15:04:01.19 & +10:26:15.78 & 0.036 & 0.024 & Sy1 & 60101023002 & A & 2015-07-14 & 23419 \\
& & & & & & 80701616002	 & B & 2022-01-09 & 53073 \\
NGC 5273 & 13:42:08.38 & +35:39:15.46 & 0.004 & 0.009 & Sy1 & 60061350002 & A & 2014-07-14 & 21117 \\
& & & & & & 90801618002	& B & 2022-07-03 & 18030 \\
NGC 0985 & 02:34:37.88 & $-$08:47:17.02 & 0.043 & 0.035 & Sy1 & 60061025002 & A & 2013-08-11 & 13895 \\
& & & & & & 60761008002 & B & 2021-09-13 & 21326 \\
HE 1143$-$1810 & 11:45:40.47 & $-$18:27:14.96 & 0.035 & 0.033 & Sy1 & 60302002002 & A & 2017-12-16 & 20960 \\
& & & & & & 60302002004 & B & 2017-12-18 & 20838 \\
& & & & & & 60302002006 & C & 2017-12-20 & 23096 \\
& & & & & & 60302002008 & D & 2017-12-22 & 20716 \\
& & & & & & 60302002010 & E & 2017-12-24 & 22378 \\
\hline
\end{tabular}
\end{table*}

\section{Sample selection and Data reduction}
\subsection{Sample selection}
We searched the archives of {\it NuSTAR} for data that are available for public
between the period November 2013 and December 2021. From this original list of objects,  we selected only sources that are Seyfert 1 type AGN. That criteria lead us to a sample of around 500 sources. Of these, to select sources with good signal to noise ratio (S/N) suitable for model fits, we selected only those sources having more than one epoch of observation and net count $\sim$ 0.1. This lead us to a final sample of 25 sources. Of this, we excluded the four sources, namely Ark 564, ESO 103-035, SWIFT J12127.4+5654 and 3C 382 for which there are reports in literature on the variation in $\rm{E_{cut}}$/$\rm{kT_{e}}$. Also, results on three other Seyfert 1 AGN, namely NGC 3227, MR 2251$-$178 and NGC 5548, were recently published by us \citep{2022A&A...662A..78P}. So, in this work, we present our analysis on the remaining 18 sources in our sample. The details of these 18 sources, along with the data available on those sources are given in Table \ref{Table-1}.

\subsection{Data reduction}
We obtained the {\it NuSTAR} data for the 18 objects, from the HEASARC archive\footnote{\href{https://heasarc.gsfc.nasa.gov/db- perl/W3Browse/w3browse.pl}{https://heasarc.gsfc.nasa.gov/db- perl/W3Browse/w3browse.pl}}.  We reduced the data in the 3$-$79 keV band using the standard {\it NuSTAR} data reduction software NuSTARDAS\footnote{\href{https://heasarc.gsfc.nasa.gov/docs/nustar/analysis/nustar swguide.pdf}{https://heasarc.gsfc.nasa.gov/docs/nustar/analysis/nustar swguide.pdf}} v1.9.3 distributed by HEASARC within HEASoft v6.26.1 Considering the passage of the satellite through the South Atlantic Anomaly (SAA) we selected SAAMODE optimized and also excluded the tentacle region. This condition was implemented based on the background reports available online. In every case, the condition that produced the maximum amount of exposure was SAACALC=2, SAAMODE=OPTIMIZED AND TENTACLE=YES. So, we chose to use this condition in {\tt nupipeline} task to take care of the background in the SAA region. The calibrated, cleaned, and screened event files were generated by running {\tt nupipeline} task using the CALDB release 20190607. To extract the source counts we chose a circular region of radius 70 arcsec centered on the source. Similarly, to extract the background counts,  we  selected a circular region of the same radius away from the source to avoid contamination from source photons. We then used the  {\tt nuproducts} task to generate energy spectra, response matrix files and auxiliary response files, for both the hard X-ray detectors housed inside the corresponding focal plane modules FPMA and FPMB. For spectral analysis, using XSPEC version 12.10.1 \citep{1996ASPC..101...17A}, we fitted the background subtracted spectra from FPMA and FPMB simultaneously (without combining them)  allowing the cross normalization factor to vary freely during spectral fits. The background subtracted source spectra were binned to have a minimum counts of 25 in each spectral bin. To get an estimate of the  model parameters that best describe the observed data, we used the chi-square ($\chi^2$) statistics and for calculating the errors in the model parameters we used the $\chi^2$ = 2.71 criterion i.e. 90 \% confidence
range in XSPEC.

\section{Spectral Analysis}
\label{sec:section3}
For all the sources in our sample, we fitted the {\it NuSTAR} data in the 3$-$79 keV band using
XSPEC. We adopted both phenomenological and physical models to fit the source spectra. For the spectral analysis we used the following four models.

\begin{enumerate}
\item const $\times$ TBabs $\times$ zTBabs $\times$ (zpo+zgauss) 
\item const $\times$ TBabs $\times$ zTBabs $\times$ (pexrav+zgauss) 
\item const $\times$ TBabs $\times$ zTBabs $\times$ (xillver/relxill/relxill+xillver)
\item const $\times$ TBabs $\times$ zTBabs $\times$ (xillverCP/relxillCP/relxillCP+xillverCP)
\end{enumerate}

In all our models that were fit to the observations, {\it const} represents the cross calibration between the two focal plane modules (FPMA and FPMB). {\it TBabs} accounts for the Milky Way hydrogen column density ($\rm N_{H}^{gal}$). The value of $\rm N_{H}^{gal}$ for each source was obtained from \cite{2013MNRAS.431..394W} and given in Table \ref{Table-1}. The host galaxy absorption ($\rm N_{H}^{INT}$) was taken care of by the {\it zTBabs} model component in which $\rm N_{H}^{INT}$ was kept free during the fit of each model. The redshift ($z$) was frozen to the values given in Table \ref{Table-1}.

The conclusion of this study is drawn based on the best fit model parameters obtained using the physically motivated Comptonization model const $\times$ TBabs $\times$ zTBabs $\times$ (xillverCP/relxillCP/relxillCP+xillverCP). Therefore, the description and the results of this particular model is discussed below while the description and findings of the other three models are discussed in Appendix \ref{sec:models}. A description of each of the sources along with the results obtained in this work is given in Appendix \ref{sec:sources}.

\begin{figure}
\vbox{
      \includegraphics[scale=0.55]{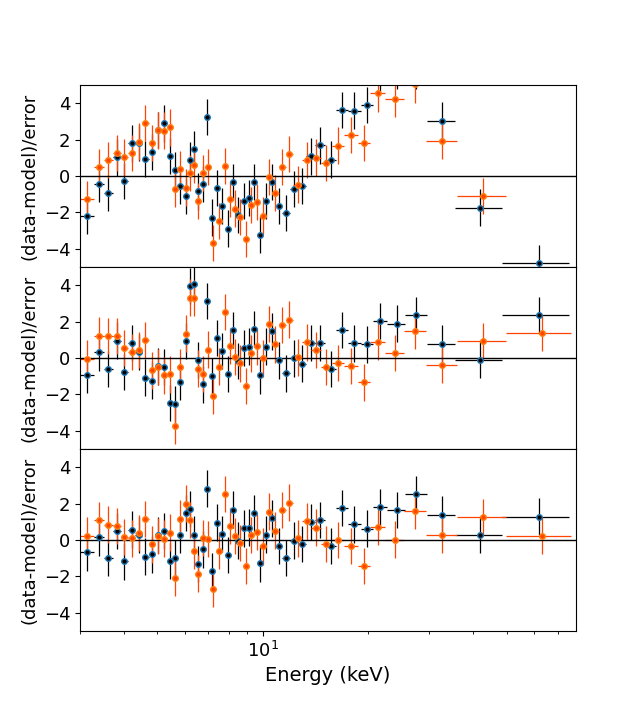}
      }
\caption{Upper panel: Ratio of data to the  model for the model fits {\it const $\times$ TBabs $\times$ ztbabs $\times$ (zpo+zgauss)}, Middle panel: {\it const $\times$ TBabs $\times$ ztbabs $\times$ (relxill)} and Bottom panel: {\it const $\times$ TBabs $\times$ ztbabs $\times$ (relxill+xillver)} to the FPMA (black dot) and FPMB (orange dot) spectra of ObsID 60501015002 (epoch B) of NGC 5506. The spectra are re-binned for visualization purposes.}\label{figure-1}  
\end{figure}

In order to estimate $\rm{kT_{e}}$ we did the spectral fitting using {\it xillverCP} \citep{2014ApJ...782...76G} and/or {\it relxillCP} \citep{2014ApJ...782...76G, 2014MNRAS.444L.100D} to take into account the distant and blurred reflection features respectively in the data. These two models are part of the {\it relxill} \citep{2014ApJ...782...76G} family and an advanced version of the disk reflection model that takes care of the physical Comptonization continuum by replacing the underlying cut-off power law in {\it xillver}/{\it relxill} with a {\it nthComp} continuum \citep{1996MNRAS.283..193Z, 1999MNRAS.309..561Z}. The model {\it xillverCP} \citep{2010ApJ...718..695G, Garc_a_2011} is a self consistent ionized reflector model that takes into account the neutral Fe K$\alpha$ ($\sim$ 6.4 keV) and Fe K$\beta$ lines ($\sim$ 7.07 keV) along with the cold reflection component present in the source spectra. Of the 18 sources in our sample, for 12 sources,  {\it xillverCP} provided  good fit statistics properly taking into account the reflection and emission line components. This model has the following form in XSPEC
\begin{equation}
    const \times TBabs \times zTBabs \times  xillverCP
\end{equation}
During the fitting we allowed the parameters $\Gamma$, $\rm{kT_{e}}$, $R$ and the normalization $N_{xillverCP}$ to vary. We considered the reflective material as  neutral and fixed the ionization parameter to 1.0 (i.e $log\xi = 0$). Also, we fixed the inclination angle ($\theta_{incl}$) to the default value of 30$^{\circ}$ and adopted solar abundance with $AF_{e}$ fixed to 1.

For four sources, we found the need for  relativistic treatment to take care of the line broadening and the relativistic smearing of the reflected radiation. Therefore for those four sources, we fitted the data with an advanced reflection model {\it relxillCP}. In XSPEC the model has the form,
\begin{equation}
    const \times TBabs \times zTBabs \times relxillCP
\end{equation}
In this model we froze a few parameters as it was not possible to constrain all the model parameters due to S/N limitations in the data. The inner ($R_{in}$) and outer radius ($R_{out}$) of the accretion disk were fixed to 1 ISCO and 400 $R_{g}$ respectively. The inclination angle ($\theta_{incl}$) was frozen to the default value of 30$^{\circ}$ and we considered a maximally spinning black hole with the spin parameter fixed to  $a_{*}$ = 0.998. We also considered a continuous disk emission profile and varied only  $\beta_{1}$, while $\beta_{2}$ was tied with $\beta_{1}$. Thus in this model fitting, the free parameters were $\Gamma$, $\rm{kT_{e}}$, $R$, $AF_{e}$, $log\xi$, $\beta_{1}$ and the normalization $N_{relxillCP}$.

In 2 out of the 13 sources in our sample, either {\it xillverCP} or {\it relxillCP} alone could not take care of the emission region completely (see Fig. \ref{figure-1}). For those two sources (NGC 3783 and NGC 5506), we considered  both the models with the following form in XSPEC,

\begin{equation}
    const \times TBabs \times zTBabs \times (relxillCP+xillverCP)
\end{equation}
In this model all the parameters that are common between {\it relxillCP} and {\it xillverCP} were tied together except $log\xi$ and the normalization. In {\it relxillCP} the accretion disk closer to the black hole was assumed to be ionized and $log\xi$ was kept free, however, within {\it xillverCP} the reflection component was assumed to be coming from more distant and neutral ($log\xi=0$) material. The best fit parameters are given in Table \ref{table-5}. The unfolded spectra with the best fit models, const $\times$ TBabs $\times$ zTBabs $\times$ (xillverCP/relxillCP/relxillCP+xillverCP) and the residues of the fit for the sample of 18 sources are given in Fig. \ref{figure-6} and \ref{figure-7}.

\begin{figure}
\hspace{-0.7 cm}
\includegraphics[scale=0.47]{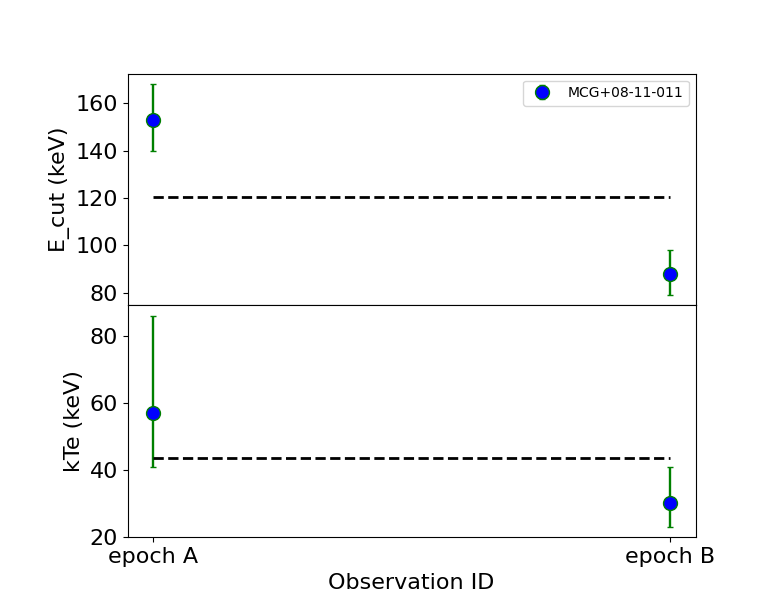}
\caption{Upper panel: Variation in $\rm{E_{cut}}$ and Lower panel: Variation in $\rm{kT_{e}}$ with the observation epochs as obtained from the {\it xillver} and {\it xillverCP} model fits respectively. The plotted errors were calculated in 90 \% confidence level. The dashed lines in each panel are fits of constant (mean of $\rm{E_{cut}}$ and $\rm{kT_{e}}$) to the data points.} \label{figure-3} \end{figure}

\section{Results and Discussion}
In this work we aimed to study the variation in the temperature of the corona in AGN. Our sample consists of 18 sources having a total of 52 epochs of observations. All these observations were analysed in a homogeneous manner. Results for few OBSIDs in a few sources are the first time measurements. However, for few sources in our sample, $\rm{E_{cut}}$ measurement were available in literature, either using data from {\it NuSTAR} in isolation or with the inclusion of data from different instruments along with that from {\it NuSTAR}. We find that our results are in general consistent (within the uncertainties) with those available in literature. For a large fraction of the objects, our analysis has provided measurement of $\rm{kT_{e}}$. 

\begin{figure}
\vbox{
      \includegraphics[scale=0.55]{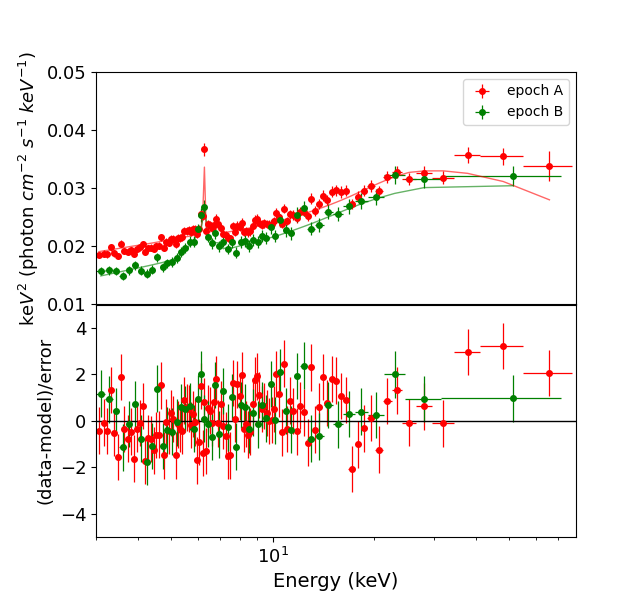}
      }
\caption{Upper panel: Unfolded spectra with the best fit model {\it const $\times$ TBabs $\times$ (xillverCP)} and Bottom panel: Residue of the to the epoch A (red dot) and epoch B (green dot) spectra of MCG+08-11-011. The spectra are re-binned for visualization purpose.}\label{figure-2}  
\end{figure}

For few sources, in this sample, there were indications of $\rm{kT_{e}}$ variation, however, due to the large error bars in the derived values, unambiguous claim on the detection of variation in $\rm{kT_{e}}$ could not be made, For example, variation in $\rm{kT_{e}}$ was detected between epoch A and epoch C of Mrk 915 (see Table \ref{table-5}), but as we could not constrain the best fit value of $\rm{kT_{e}}$ in epoch A, unambiguous claim on the detection of variation in $\rm{kT_{e}}$ in the source could not be made. For only one source namely MCG+08+11-011, we found variation in $\rm{kT_{e}}$ between two epochs separated by about five years. However, the variability measured in $\rm{kT_{e}}$ between epoch A and epoch B of MCG+08-11-011 is statistically marginal, the measurements being mutually consistent within the 90 \% confidence level. For this source the variation in $\rm{E_{cut}}$ could also be ascertained from both the {\it pexrav} and {\it xillver} model fit to the source spectra. The variation of $\rm{E_{cut}}$ and $\rm{kT_{e}}$ as obtained from the physical model fit ({\it xillver} and {\it xillverCP}) with the epochs of observations is plotted in Fig. \ref{figure-3} in which the dashed lines in each panel represent the mean value of these two parameters respectively. To understand the reasons behind the change in $\rm{kT_{e}}$, we examined the correlations between $\Gamma$ against flux, $\rm{kT_{e}}$ against flux, $R$ against flux and $\rm{kT_{e}}$ against the optical depth ($\tau$) of the corona. We calculated $\tau$ using the following equation (\citep{1996MNRAS.283..193Z, 1999MNRAS.309..561Z}) 

\begin{equation*}
    \tau = \sqrt{\frac{9}{4} + \frac{3}{\theta\Big[\Big(\Gamma + \frac{1}{2}\Big)^2 - \frac{9}{4}\Big]}} - \frac{3}{2}  \\
\end{equation*}   

where $\theta = {kT_e}/{m_{e}c^2}$. We also estimated the flux for const $\times$ TBabs $\times$ zTBabs $\times$ (xillverCP/relxillCP/relxillCP+xillverCP) in the 3$-$79 keV band using the XSPEC model {\it cflux}. The values of flux as well as $\tau$ are given in Table \ref{table-6}. The corresponding errors in flux and $\tau$ were calculated at the 90\%  confidence level.  The unfolded epoch A and epoch B FPMA spectra with the best fit model, const $\times$ TBabs $\times$ (xillverCP) and the residues of the fit are given in Fig. \ref{figure-2}. 

\begin{figure}
\hspace{-0.7 cm}
\includegraphics[scale=0.47]{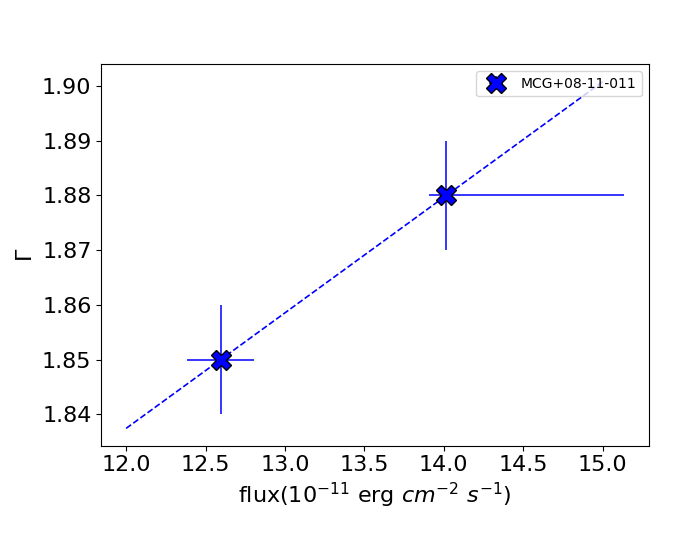}
\caption{The relation between $\Gamma$ and flux. The dashed line is the linear least squares fit to the data.} \label{figure-4} 
\end{figure}

\begin{figure*}
\hbox{
     \includegraphics[scale=0.33]{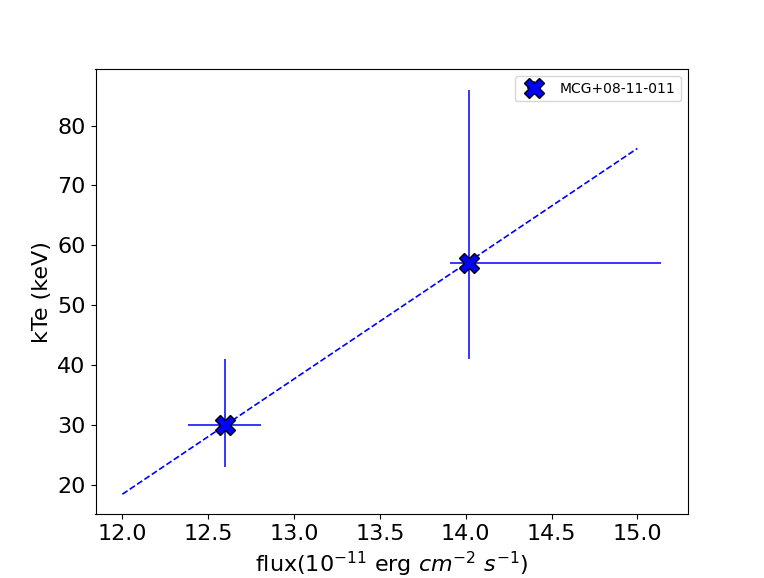}
     \includegraphics[scale=0.33]{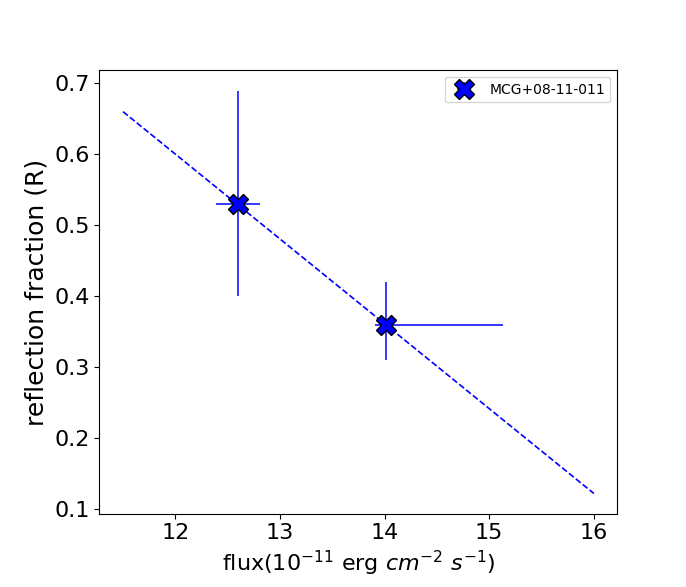}
     \includegraphics[scale=0.33]{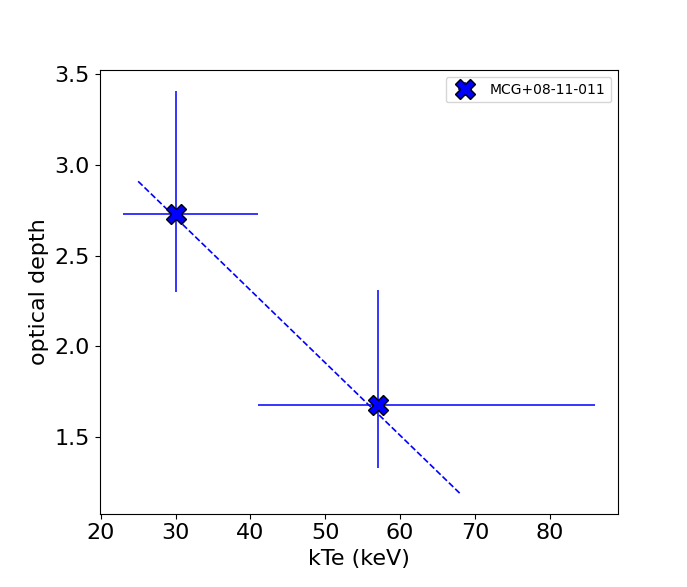}
     }
\caption{Left panel: The correlation between $\rm{kT_{e}}$ and flux. Middle panel: The relation between $R$ and flux. Right panel: The relation between $\tau$ and $\rm{kT_{e}}$.} The dashed lines in all the panels are the linear least squares fit to the data. \label{figure-5}       
\end{figure*}

Like all other bright Seyfert galaxies, MCG+08-11-011 also became softer when it brightened in X-ray \citep{2003ApJ...593...96M, 2014A&A...563A..57S, 2014ApJ...794...62B}. In Fig. \ref{figure-4} the {\it softer-when-brighter} trend is shown for this source. We also found MCG+08-11-011 in its higher flux state (14.02$^{+1.12}_{-0.11}$ $\times$ $10^{-11} erg cm^{-2} s^{-1}$) on August 2016 (epoch A) with a hotter corona of $\rm{kT_{e}}$ $=$ 57$^{+29}_{-16}$ keV. On December 2021 (epoch B) the source was found to have a cooler corona ($\rm{kT_{e}}$ $=$ 30$^{+11}_{-7}$ keV) with a lower flux of 12.60$^{+0.21}_{-0.21}$ $\times$ $10^{-11} erg cm^{-2} s^{-1}$ (see Fig. \ref{figure-2} and left panel of Fig. \ref{figure-5}). The source was optically thinner on August 2016 with a $\tau$ $=$ 1.68$^{+0.63}_{-0.35}$ and $\tau$ increased to 2.73$^{+0.68}_{-0.43}$ on December 2021. The correlations between $\rm{kT_{e}}$ against flux (left panel) and $\rm{kT_{e}}$ against $\tau$ (right panel) is shown in Fig. \ref{figure-5}. Thus, in MCG+08-11-011 we found a  {\it hotter-when-brighter and softer-when-brighter} behaviour. In the brightest epoch the corona is also found to be optically thinner. The {\it hotter-when-brighter and softer-when-brighter} behaviour could be explained by the changing nature of the corona. All such correlations point to the change in the geometry of the corona that in turn results the variation in the heating/cooling of the AGN corona. The possible scenario that can explain such variation is that at the low flux state the corona lies closer to the black hole and remains optically thick. In this position, the fraction of the reflected emission to the primary emission increases thereby increasing the reflection fraction $R$. As the source brightens, the corona moves vertically away from the central engine. This vertical movement results in an optically thinner and hotter corona with lesser amount of reflected emission off the accretion disk, thus a softer X-ray spectrum is expected. The observed correlations among $R$, flux and $\tau$ with $\rm{kT_{e}}$ for MCG+08-11-011, shown in Fig. \ref{figure-5} are evidences for the {\it vertically out flowing corona} \citep{2016MNRAS.456.2722K, 2018ApJ...863...71Z, 2021ApJ...921...46B} as a cause of the observed $\rm{kT_{e}}$ variation in this source.

Recently, \cite{2021MNRAS.tmp...68K} found a $\Lambda$ shaped correlation between $\rm{E_{cut}}$ and $\Gamma$ for SWIFT J2127.4+5654. The authors argued for both the {\it Compton cooling} and {\it vertically out flowing corona} scenario responsible for the observed $\rm{E_{cut}}$ variation in this source. The authors also claimed to have found a break point in $\Gamma$ ($\gtrsim$ 2.05) after which the {\it Compton cooling} would dominate over the {\it vertically out flowing corona} scenario.  \cite{2021ApJ...921...46B} reported variation of $\rm{kT_{e}}$ in a Seyfert 2 galaxy ESO 103-035, wherein they found a positive correlation between $\rm{kT_{e}}$  and flux. According to the authors, the {\it hotter-when-brighter} trend could be due to the AGN variability being driven by coronal heating variation or due to the AGN variability driven by the changes in the seed photon flux. Previously in Ark 564 the same authors \citep{2020MNRAS.492.3041B} found evidence of $\rm{kT_{e}}$ variation with  {\it cooler-when-brighter} nature. To explain the opposite behaviour of the corona in the two sources namely ESO 103-035 and Ark 564 the authors pointed to a cut off in $\Gamma$. From Comptonization model fits to the data,  they found $\Gamma$ $<$ 2.00 in ESO 103-035 and $\Gamma$ $>$ 2.00 in Ark 564.  \cite{2021ApJ...921...46B} suggested that the difference in $\Gamma$ between the two sources is the reason behind the two completely different behaviour observed in them. In the case of MCG+08-11-011 too, we found $\Gamma$ $<$ 2, which could be the driving factor for the {\it hotter-when-brighter} behaviour observed. We also noticed an anti-correlation between $\tau$ and $\rm{kT_{e}}$ (see right panel of Fig. \ref{figure-5}). According to \cite{2018A&A...614A..37T} the observed negative correlations among these two parameters in most of the Seyferts is the proof of the disturbance in the fixed disk-corona model in radiative balance. This kind of anti-correlation between $\tau$ and $\rm{kT_{e}}$ can take place if either there is a change in the ratio between the intrinsic disk emission to the total disk emission or there is a change in geometry/position of the corona itself. The observed correlation between various physical parameters in MCG+08-11-011, tend to favor a change in the geometry or position of the corona between the two epochs leading to the observed $\rm{kT_{e}}$ variations. The idea that sources with positive correlation between $\rm{kT_{e}}$ and flux have $\Gamma$ $<$ 2 and those with $\Gamma$ $>$ 2 show negative correlation between $\rm{kT_{e}}$ and flux needs to be confirmed, which necessitates detection of $\rm{kT_{e}}$ variation and its correlation with flux, in many sources.

\begin{table*}
\caption{Best fit parameters of the  Model {\it const $\times$ TBabs $\times$ zTBabs $\times$ (xillverCP/relxillCP/(relxillCP+xillverCP))} to the source spectra. The model normalization is in units of photons keV$^{-1}$ cm$^{-2}$s$^{-1}$, and $\rm N_{H}^{INT}$ is the host galaxy hydrogen column density in units of atoms $\rm{cm^{-2}}$. The asterisk(*) against entries indicates that they are frozen.}\label{table-5}
\centering
\begin{tabular}{p{0.11\linewidth}p{0.03\linewidth}p{0.06\linewidth}p{0.06\linewidth}p{0.06\linewidth}p{0.06\linewidth}p{0.06\linewidth}p{0.06\linewidth}p{0.06\linewidth}p{0.06\linewidth}p{0.06\linewidth}p{0.06\linewidth}}
\hline
Source & Epoch & $\rm N_{H}^{INT}$ & $\Gamma$ & $\rm{kT_{e}}$ & R & $log\xi$ & $\rm AF_{e}$ & $\beta_{1}$ & $\rm N_{xillverCP}$ & $\rm N_{relxillCP}$ & $\chi^2/dof$   \\
& & & & & & & & & & & \\
& & ($10^{22}$) & & (keV) & & & & &  ($10^{-4}$) & ($10^{-4}$) & \\
\hline
1H 0419$-$577 & A & 2.61$^{+0.68}_{-0.66}$ & 1.84$^{+0.02}_{-0.02}$ & 18$^{+5}_{-3}$ & 0.14$^{+0.11}_{-0.10}$ & - & - & - & 0.66$^{+0.03}_{-0.01}$ & - &  736/733 \\
& B & 2.34$^{+0.72}_{-0.71}$ & 1.82$^{+0.02}_{-0.02}$ & 19$^{+10}_{-4}$ & $<$0.17 & - & - & - & 0.78$^{+0.01}_{-0.01}$ & - &  659/680 \\
& C & 1.89$^{+0.36}_{-0.36}$ & 1.82$^{+0.01}_{-0.01}$ & 15$^{+1}_{-1}$ & 0.16$^{+0.06}_{-0.06}$ & - & - & - & 0.79$^{+0.07}_{-0.06}$ & - & 1180/1156 \\
Mrk 915 & A & 5.06$^{+0.71}_{-0.71}$ & 1.88$^{+0.02}_{-0.01}$ & $>$77 & 0.46$^{+0.17}_{-0.15}$ & - & - & - & 0.62$^{+0.10}_{-0.02}$ & - & 655/650 \\
& B & 6.31$^{+0.94}_{-0.92}$ & 1.82$^{+0.02}_{-0.01}$ & 30$^{+124}_{-12}$ & 0.39$^{+0.22}_{-0.17}$ & - & - & - & 0.36$^{+0.01}_{-0.01}$ & - & 549/543 \\
& C & 8.00$^{+1.10}_{-1.16}$ & 1.90$^{+0.03}_{-0.03}$ & 17$^{+7}_{-3}$ & 0.89$^{+0.53}_{-0.24}$ & - & - & - & 0.25$^{+0.01}_{-0.01}$ & - & 423/435 \\
3C 111 & A & 2.43$^{+0.60}_{-0.55}$ & 1.87$^{+0.02}_{-0.02}$ & $>$63 & 0.12$^{+0.10}_{-0.09}$ & - & - & - & 3.13$^{+0.51}_{-0.05}$ & - &  669/683 \\
& B & 2.42$^{+0.38}_{-0.39}$ & 1.84$^{+0.01}_{-0.01}$ & 40$^{+47}_{-12}$ & 0.08$^{+0.07}_{-0.06}$ & - & - & - & 2.50$^{+0.02}_{-0.02}$ & - &  981/949 \\
& C & 2.93$^{+0.45}_{-0.45}$ & 1.77$^{+0.02}_{-0.01}$ & $>$50 & $<$0.13 & - & - & - & 2.29$^{+0.51}_{-0.03}$ & - & 962/899 \\
NGC 3783 & A & - & 1.81$^{+0.05}_{-0.04}$ & 65$^{+107}_{-24}$ & 0.84$^{+0.33}_{-0.25}$ & 2.75$^{+0.16}_{-0.08}$ & $1.00^{*}$ & 2.67$^{+0.36}_{-0.32}$ & 0.62$^{+0.39}_{-0.47}$ & 2.01$^{+0.20}_{-0.18}$ & 1108/1038 \\
& B & 3.71$^{+0.85}_{-0.78}$ & 1.87$^{+0.04}_{-0.04}$ & $>$100 & 1.16$^{+0.46}_{-0.36}$ & 2.88$^{+0.24}_{-0.24}$   & $1.00^{*}$ & 3.61$^{+0.76}_{-0.60}$ & 1.61$^{+0.40}_{-0.48}$ & 1.59$^{+0.23}_{-0.23}$ & 1037/1002 \\
& C & 8.88$^{+1.53}_{-1.30}$ & 1.90$^{+0.05}_{-0.04}$ & $>$99 & 1.22$^{+0.79}_{-0.52}$ & 3.00$^{+0.35}_{-0.32}$ & $1.00^{*}$ & 3.56$^{+0.71}_{-0.72}$ & 2.16$^{+0.47}_{-0.60}$ & 1.44$^{+0.27}_{-0.34}$ & 940/792 \\
& D & 6.73$^{+1.10}_{-1.07}$ & 1.88$^{+0.05}_{-0.05}$ & $>$95 & 1.38$^{+0.71}_{-0.51}$ & 2.97$^{+0.24}_{-0.26}$ & $1.00^{*}$ & 3.60$^{+0.56}_{-0.59}$ & 1.92$^{+0.49}_{-0.57}$ &  1.64$^{+0.29}_{-0.33}$ & 1052/913 \\
NGC 7469 & A & - & 2.00$^{+0.02}_{-0.02}$ & 43$^{+42}_{-14}$ & 0.75$^{+0.19}_{-0.18}$ & - & - & - & 1.58$^{+0.03}_{-0.03}$ & - & 615/634 \\
& B & - & 1.91$^{+0.02}_{-0.02}$ & $>$45 & 0.51$^{+0.18}_{-0.15}$ & - & - & - & 1.82$^{+0.04}_{-0.03}$ & - & 624/630 \\
& C & - & 1.98$^{+0.02}_{-0.02}$ & $>$38 & 0.93$^{+0.23}_{-0.21}$ & - & - & - & 1.34$^{+0.03}_{-0.03}$ & - & 565/607 \\
& D & - & 1.94$^{+0.02}_{-0.02}$ & $>$84 & 0.55$^{+0.15}_{-0.14}$ & - & - & - & 1.64$^{+0.03}_{-0.03}$ & - & 627/635 \\
& E & - & 1.92$^{+0.02}_{-0.02}$ & $>$60 & 0.60$^{+0.19}_{-0.17}$ & - & - & - & 1.74$^{+0.06}_{-0.03}$ & - & 674/620 \\
& F & - & 1.89$^{+0.02}_{-0.02}$ & $>$81 & 0.56$^{+0.19}_{-0.16}$ & - & - & - & 1.67$^{+0.05}_{-0.04}$ & - & 601/606 \\
& G & - & 1.90$^{+0.02}_{-0.02}$ & $>$53 & 0.45$^{+0.15}_{-0.13}$ & - & - & - & 1.99$^{+0.05}_{-0.03}$ & - & 682/696 \\
Mrk 110 & A & - & 1.81$^{+0.01}_{-0.01}$ & 23$^{+3}_{-2}$ & 0.10$^{+0.03}_{-0.03}$ & $3.00^{*}$ & $>$8.94 & $3.0^{*}$ & - & 1.98$^{+0.05}_{-0.05}$ & 1671/1475 \\
& B & - & 1.81$^{+0.02}_{-0.02}$ & 22$^{+7}_{-4}$ & 0.16$^{+0.08}_{-0.07}$ & 3.10$^{+0.30}_{-0.24}$ & $>$7.85 & $3.0^{*}$ & - & 1.37$^{+0.08}_{-0.08}$ & 1041/1025 \\
& C & - & 1.77$^{+0.01}_{-0.03}$ & 25$^{+9}_{-5}$ & 0.17$^{+0.07}_{-0.08}$ & 3.04$^{+0.33}_{-0.20}$ & $>$7.76 & $3.0^{*}$ & - & 1.20$^{+0.08}_{-0.08}$ & 1061/1006 \\
UGC 06728 & A & - & 1.73$^{+0.07}_{-0.07}$ & 17$^{+34}_{-5}$ & 0.33$^{+0.41}_{-0.22}$ & $<$3.34  & $1.00^{*}$ & $<$3.98 & - & 0.32$^{+0.49}_{-0.43}$ & 188/197 \\
& B & - & 1.70$^{+0.07}_{-0.06}$ & $>$23 & 0.42$^{+1.17}_{-0.17}$ & 3.11$^{+0.58}_{-0.75}$  & $1.00^{*}$ & 2.30$^{+0.47}_{-0.97}$ & - & 0.67$^{+0.20}_{-0.38}$ & 720/707 \\
NGC 4258 & A & 11.73$^{+1.69}_{-1.40}$ & 1.81$^{+0.03}_{-0.03}$ & 13$^{+7}_{-3}$ & $<$0.51 & - & - & - & 0.20$^{+0.18}_{-0.09}$ & - & 256/277 \\
& B & 13.08$^{+1.20}_{-1.15}$ & 1.89$^{+0.04}_{-0.03}$ & 9$^{+2}_{-1}$ & $<$0.93 & - & - & - & 0.15$^{+0.06}_{-0.05}$ & - &  438/408 \\
KUG 1141+371 & A & 1.64$^{+0.79}_{-0.79}$ & 1.97$^{+0.03}_{-0.03}$ & $>$22 & 0.37$^{+0.21}_{-0.18}$ & - & - & - & 0.64$^{+0.16}_{-0.14}$ & - & 430/485 \\
& B & 2.97$^{+1.25}_{-1.38}$ & 1.84$^{+0.04}_{-0.05}$ & $>$16 & $<$0.44 & - & - & - & 0.55$^{+0.29}_{-0.37}$ & - & 239/247 \\
MCG-06-30-15 & A & - & 2.00$^{+0.08}_{-0.07}$ & $>$50 & 1.69$^{+2.13}_{-0.47}$ & 3.10$^{+0.17}_{-0.19}$ & 1.01$^{+1.47}_{-0.27}$ & 2.86$^{+0.29}_{-0.21}$ & - & 1.43$^{+0.43}_{-0.63}$ & 548/732 \\
& B & - & 2.02$^{+0.03}_{-0.03}$ & $>$74 & 1.03$^{+0.15}_{-0.15}$ & 2.84$^{+0.09}_{-0.09}$ & 3.34$^{+0.89}_{-0.64}$ & 2.94$^{+0.15}_{-0.13}$ & - & 2.18$^{+0.10}_{-0.11}$ & 1444/1371 \\
& C & - & 1.88$^{+0.07}_{-0.04}$ & 28$^{+59}_{-9}$ & 1.12$^{+0.45}_{-0.22}$ & $<$2.67 & $>$3.05 & 2.69$^{+0.35}_{-0.47}$ & - & 1.39$^{+0.08}_{-0.08}$ & 707/740 \\
NGC 5506 & A & 4.03$^{+0.54}_{-0.97}$ & 1.83$^{+0.04}_{-0.08}$ & $>$42 & 0.27$^{+0.13}_{-0.12}$ & 3.11$^{+0.35}_{-0.24}$ & 1.39$^{+0.72}_{-0.37}$ & 2.08$^{+0.57}_{-0.61}$ & 1.88$^{+0.64}_{-0.64}$ & 3.14$^{+0.15}_{-0.52}$ & 1239/1228 \\
& B & 2.50$^{+0.68}_{-0.63}$ & 1.75$^{+0.07}_{-0.03}$ & 27$^{+25}_{-5}$ & 0.63$^{+0.12}_{-0.09}$ & 2.43$^{+0.35}_{-0.34}$ & 4.97$^{+2.01}_{-1.36}$ & 2.78$^{+0.30}_{-0.28}$ & 0.58$^{+0.36}_{-0.14}$ & 2.69$^{+0.15}_{-0.16}$ & 1254/1310 \\
& C & 1.96$^{+0.59}_{-0.73}$ & 1.71$^{+0.05}_{-0.05}$ & 25$^{+19}_{-6}$ & 0.50$^{+0.14}_{-0.08}$ & 3.00$^{+0.09}_{-0.17}$ & 4.61$^{+4.80}_{-1.48}$ & 2.97$^{+0.22}_{-0.24}$ & 0.61$^{+0.36}_{-0.24}$ & 3.54$^{+0.21}_{-0.18}$ & 1218/1293 \\
MCG+08-11-011 & A & - & 1.88$^{+0.01}_{-0.01}$  & 57$^{+29}_{-16}$  & 0.36$^{+0.06}_{-0.05}$  & - & - & - & 2.92$^{+0.02}_{-0.02}$  & - & 1408/1316 \\
& B & - & 1.85$^{+0.01}_{-0.01}$  & 30$^{+11}_{-7}$  & 0.53$^{+0.16}_{-0.13}$  & - & - & - & 2.30$^{+0.03}_{-0.07}$  & - & 804/745 \\
GRS 1734-292 & A & 2.58$^{+0.45}_{-0.44}$ & 1.83$^{+0.01}_{-0.01}$ & 17$^{+2}_{-2}$ & 0.19$^{+0.09}_{-0.08}$ & - & - & - & 3.72$^{+0.08}_{-0.05}$ & - & 948/865 \\
& B & 3.75$^{+0.46}_{-0.44}$ & 1.81$^{+0.01}_{-0.01}$ & 20$^{+4}_{-2}$ & 0.24$^{+0.08}_{-0.09}$ & - & - & - & 2.98$^{+0.05}_{-0.04}$ & - & 818/856 \\
Mrk 926 & A & - & 1.80$^{+0.01}_{-0.01}$ & 45$^{+39}_{-9}$ & $<$0.12 & - & - & - &  3.27$^{+0.02}_{-0.02}$ & - & 1558/1468 \\ 
& B & - & 1.77$^{+0.03}_{-0.03}$ & $>$20 & $<$0.35 & - & - & - &  1.19$^{+0.03}_{-0.03}$ & - & 478/494 \\
Mrk 841 & A & - & 1.86$^{+0.03}_{-0.02}$ & $>$20 & 0.35$^{+0.23}_{-0.18}$ & - & - & - &  0.91$^{+0.02}_{-0.02}$ & - & 479/507 \\
& B & - & 1.88$^{+0.02}_{-0.02}$ & 33$^{+22}_{-11}$ & 0.52$^{+0.17}_{-0.15}$ & - & - & - &  0.80$^{+0.01}_{-0.01}$ & - & 741/744 \\
NGC 5273 & A & - & 1.67$^{+0.03}_{-0.03}$ & 17$^{+5}_{-3}$ & 0.83$^{+0.31}_{-0.25}$ & $2.00^{*}$ & $>$7.94 & 2.59$^{+0.29}_{-0.34}$ & - & 0.86$^{+0.04}_{-0.06}$  & 606/574 \\ 
& B & - & 1.55$^{+0.03}_{-0.03}$ & 15$^{+3}_{-3}$ & 0.76$^{+0.51}_{-0.34}$ & $2.00^{*}$ & 2.63$^{+2.20}_{-1.39}$ & 2.80$^{+0.36}_{-0.40}$ & - & 0.68$^{+0.12}_{-0.12}$  & 487/506 \\
NGC 0985 & A & 7.65$^{+1.51}_{-1.58}$ & 1.93$^{+0.04}_{-0.04}$ & $>$23 & 0.58$^{+0.39}_{-0.30}$ & - & - & - &  0.65$^{+0.02}_{-0.02}$ & - & 302/267 \\
& B & - & 1.89$^{+0.03}_{-0.03}$ & $>$28 & 0.51$^{+0.27}_{-0.23}$ & - & - & - &  0.86$^{+0.02}_{-0.02}$ & - & 391/466 \\
HE 1143$-$1810 & A & 1.87$^{+0.78}_{-0.77}$ & 1.86$^{+0.02}_{-0.02}$ & 20$^{+13}_{-4}$ & 0.20$^{+0.17}_{-0.15}$ & - & - & - &  1.11$^{+0.02}_{-0.03}$ & - & 491/506 \\
& B & 3.25$^{+0.84}_{-0.82}$ & 1.92$^{+0.03}_{-0.03}$ & $>$28 & 0.39$^{+0.21}_{-0.18}$ & - & - & - &  1.23$^{+0.13}_{-0.03}$ & - & 470/486 \\
& C & - & 1.85$^{+0.02}_{-0.02}$ & 27$^{+19}_{-7}$ & 0.26$^{+0.15}_{-0.13}$ & - & - & - &  1.37$^{+0.02}_{-0.02}$ & - & 534/597 \\
& D & 2.26$^{+0.71}_{-0.68}$ & 1.91$^{+0.03}_{-0.02}$ & $>$28 & 0.25$^{+0.16}_{-0.14}$ & - & - & - &  1.52$^{+0.15}_{-0.04}$ & - & 591/561 \\
& E & 1.82$^{+0.68}_{-0.68}$ & 1.88$^{+0.02}_{-0.02}$ & 37$^{+269}_{-15}$ & 0.18$^{+0.15}_{-0.12}$ & - & - & - &  1.42$^{+0.01}_{-0.03}$ & - & 590/568 \\
\hline\hline
\end{tabular}
\end{table*}

\begin{table}
\caption{Calculated flux and $\tau$.} \label{table-6}
\centering
\begin{tabular}{llcc}
\hline
Source & Epoch & Flux & $\tau$  \\
&& ($10^{-11} erg cm^{-2} s^{-1}$) & \\
\hline
MCG-08-11-011 & A & 14.02$^{+1.12}_{-0.12}$ & 1.68$^{+0.63}_{-0.35}$ \\
& B & 12.60$^{+0.21}_{-0.21}$ & 2.73$^{+0.68}_{-0.43}$ \\
\hline
\end{tabular}
\end{table}

\section{Summary}
In this work we have carried out phenomenological and physical model fits 
to 52 epochs of data from {\it NuSTAR} on 18 sources. These were carried
out to extract important spectral parameters of the sources such as
$\Gamma$, $\rm{kT_{e}}$, reflection fraction etc. and investigate for 
variation in $\rm{kT_{e}}$ in these sources. The results of this work are 
summarized below
\begin{enumerate}
\item From Comptonization model fits to the data, we found that the
S/N in the data is sufficient to constrain $\rm{kT_{e}}$ and $\rm{E_{cut}}$
in most of the sources. For few epochs of observation we obtained 
only lower limits. 
\item On comparing our results on few sources, to their earlier results
in literature, we found that our results are in general consistent with
that in the literature.
\item For few sources, from spectral analysis we found variation in $\rm{E_{cut}}$, however, variation of  $\rm{kT_{e}}$ in them could not be confirmed except in MCG+08-11-011.
\item For MCG +08-11-011, we found variation of $\rm{kT_{e}}$ at the 90\% confidence between two epochs separated by about five years. We also found the source to show a {\it hotter-when-brighter and softer-when-brighter} behaviour. Our observations tend to favour
a scenario of  change in the position or geometry of the corona leading to 
variation in the measured $\rm{kT_{e}}$ values.
\end{enumerate}

With this study the number of sources with known $\rm{kT_{e}}$ variation, has increased to 11. Among the sources that show $\rm{kT_{e}}$, variation, both
{\it hotter-when-brighter} and {\it cooler-when-brighter} trend were observed.
Thus the observed relation between the temperature of the corona and the
source brightness is found to be different among sources. Therefore,  it is imperative to find more sources that show variation in the temperature of the corona. Such a data set will enable us to  probe better the correlation of $\rm{kT_{e}}$ with various physical properties and thereby provide the needed inputs to enhance our understanding of AGN corona. 

\section*{Acknowledgements}
We thank the anonymous referee for his/her useful comments and suggestions which improved the quality and the clarity of the paper. We thank the {\it NuSTAR} Operations, Software and Calibration teams for support with the execution and analysis of these observations. This research has made use of the {\it NuSTAR} Data Analysis Software (NuSTARDAS) jointly developed by the ASI Science Data Center (ASDC, Italy) and the California Institute of Technology (USA). This research has made use of data and/or software provided by the High Energy Astrophysics Science Archive Research Center (HEASARC), which is a service of the Astrophysics Science Division at NASA/GSFC.

\section*{Data Availability}
All data used in this work are publicly available in the {\it NuSTAR} (\url{https://heasarc.gsfc.nasa.gov/docs/nustar/nustar_archive.html}) science archive.



\bibliographystyle{mnras}
\bibliography{example} 



\appendix
\section{Best fit unfolded spectra}
\label{sec:bestfit}
\begin{figure*}
\vbox{
     \hspace{-0.05cm}\includegraphics[scale=0.27]{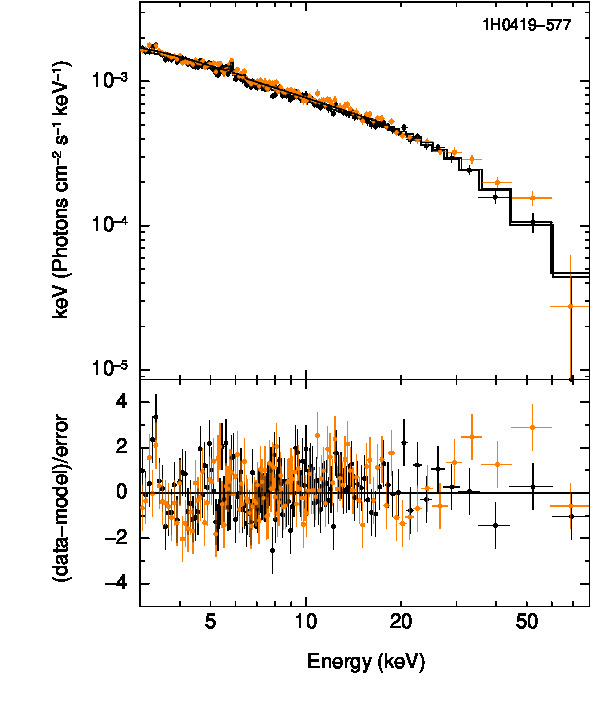}
     \hspace{-0.05cm}\includegraphics[scale=0.27]{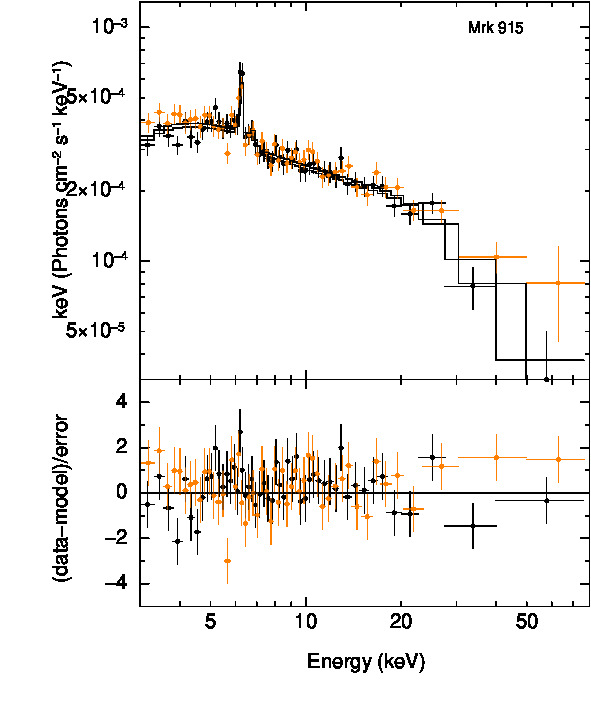}
     \hspace{-0.05cm}\includegraphics[scale=0.27]{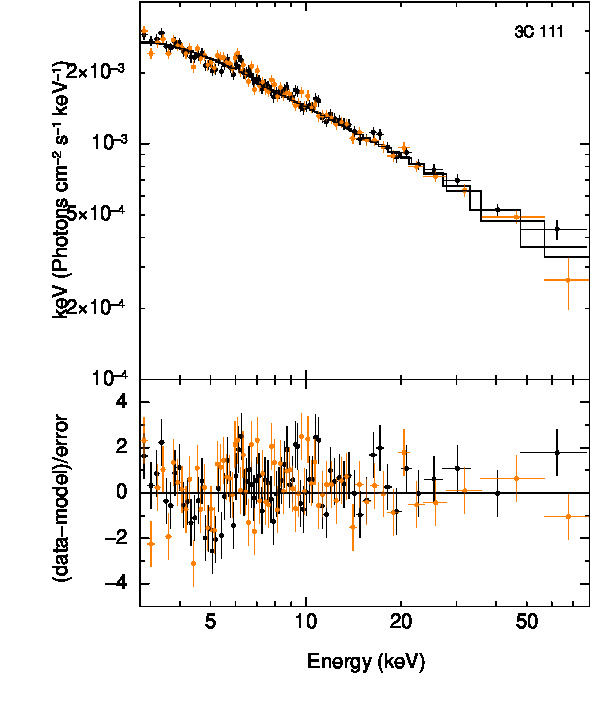}
     }
\vspace{-0.5cm}
\end{figure*}
\begin{figure*}
\vbox{
     \hspace{-0.05cm}\includegraphics[scale=0.27]{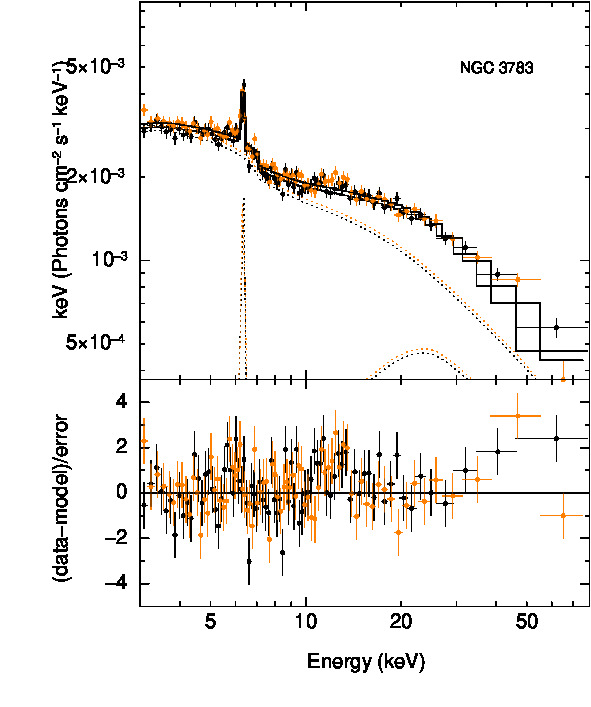}
     \hspace{-0.05cm}\includegraphics[scale=0.27]{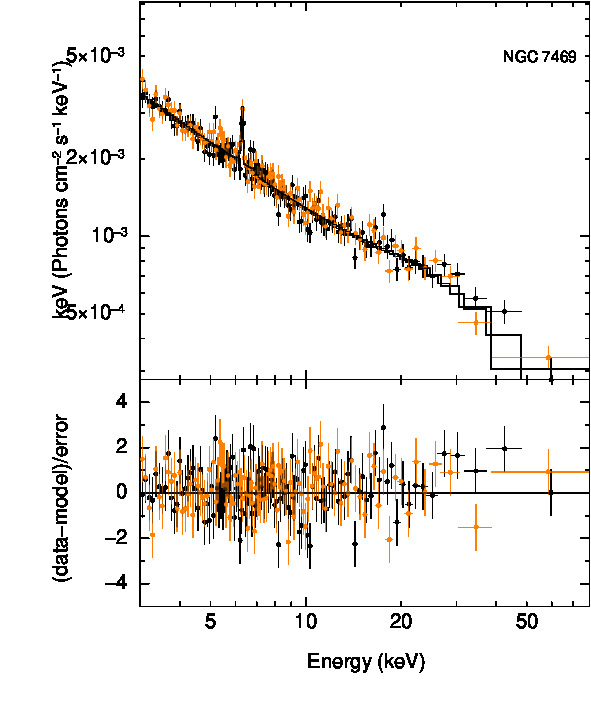}
     \hspace{-0.05cm}\includegraphics[scale=0.27]{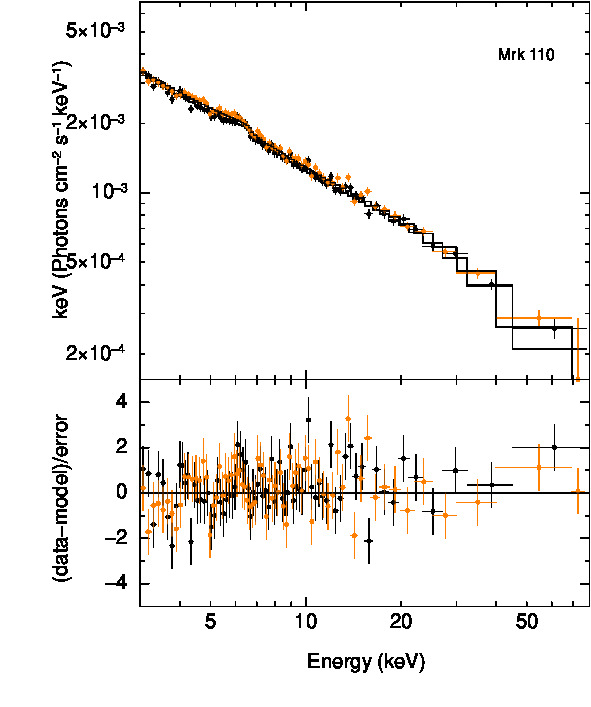}
     }
\vspace{-0.5cm}
\end{figure*}

\begin{figure*}
\vbox{
     \hspace{-0.05cm}\includegraphics[scale=0.27]{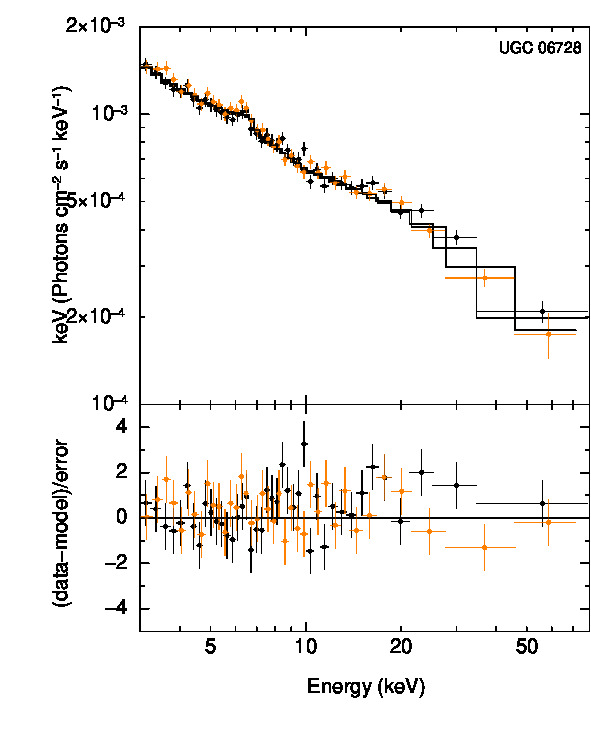}
     \hspace{-0.05cm}\includegraphics[scale=0.27]{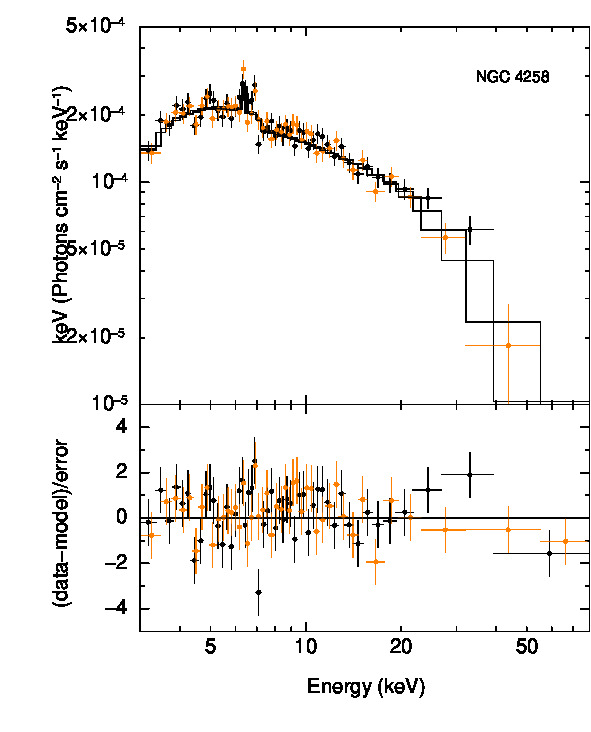}
     \hspace{-0.05cm}\includegraphics[scale=0.27]{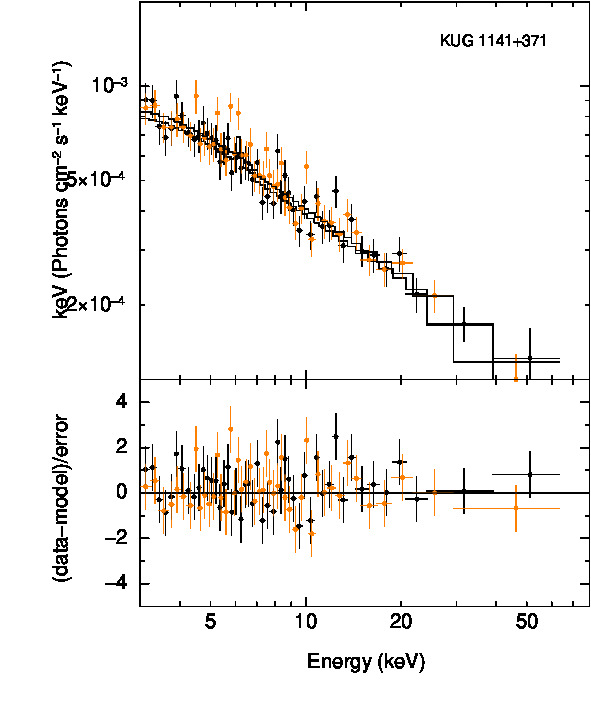}
     }
\vspace{-0.5cm}
\caption{Upper panel: The unfolded spectra and the best fit models ({\it const $\times$ TBabs $\times$ ztbabs(xillverCP/relxillCP/relxillCP+xillverCP))}) along with the residuals (lower panel) of the spectral fitting plots to the sample. Here black points are for FPMA and orange points are for FPMB. The spectra are re-binned for visualization purposes only.}
\label{figure-6}
\end{figure*}

\begin{figure*}
\vbox{
     \hspace{-0.05cm}\includegraphics[scale=0.27]{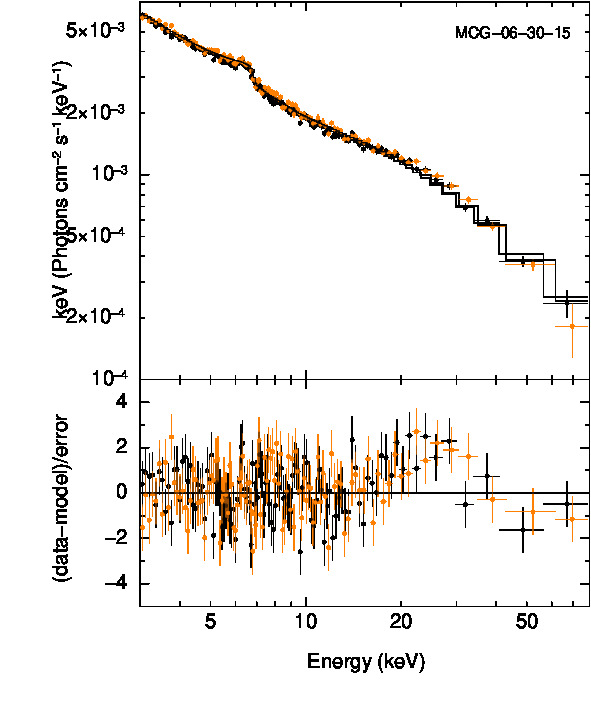}
     \hspace{-0.05cm}\includegraphics[scale=0.27]{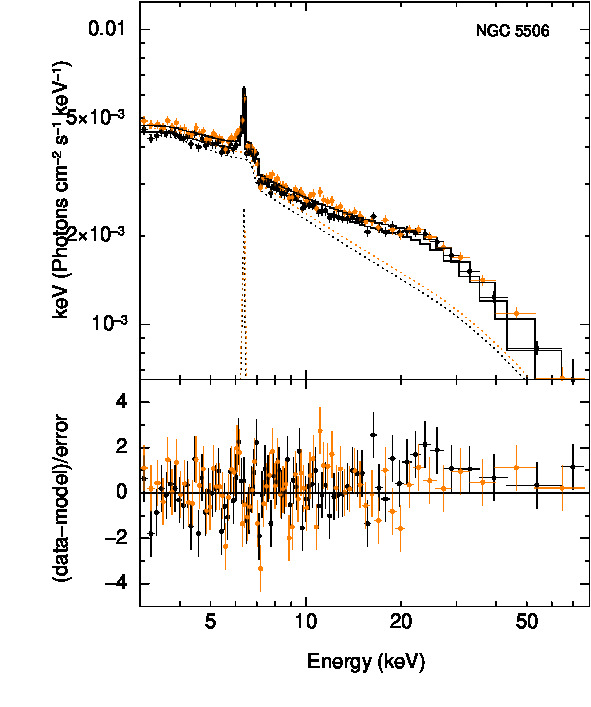}
     \hspace{-0.05cm}\includegraphics[scale=0.27]{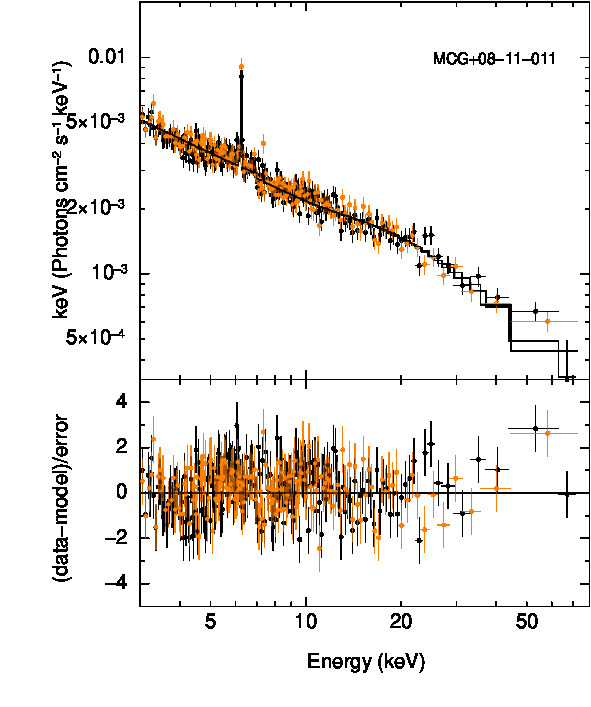}
     }
\vspace{-0.5cm}
\end{figure*}

\begin{figure*}
\vbox{
     \hspace{-0.05cm}\includegraphics[scale=0.27]{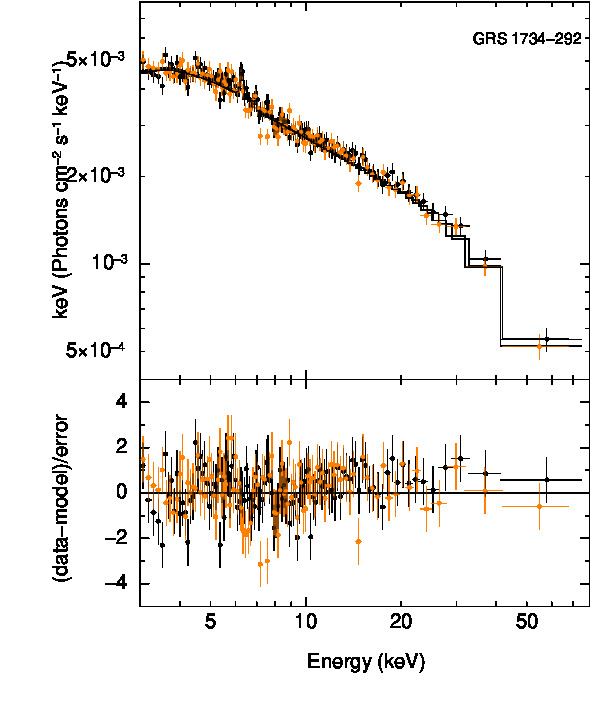}
     \hspace{-0.05cm}\includegraphics[scale=0.27]{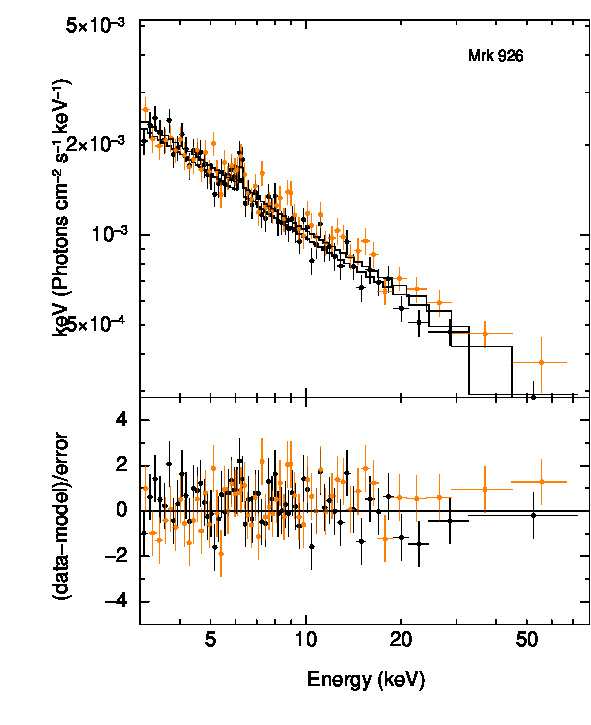}
     \hspace{-0.05cm}\includegraphics[scale=0.27]{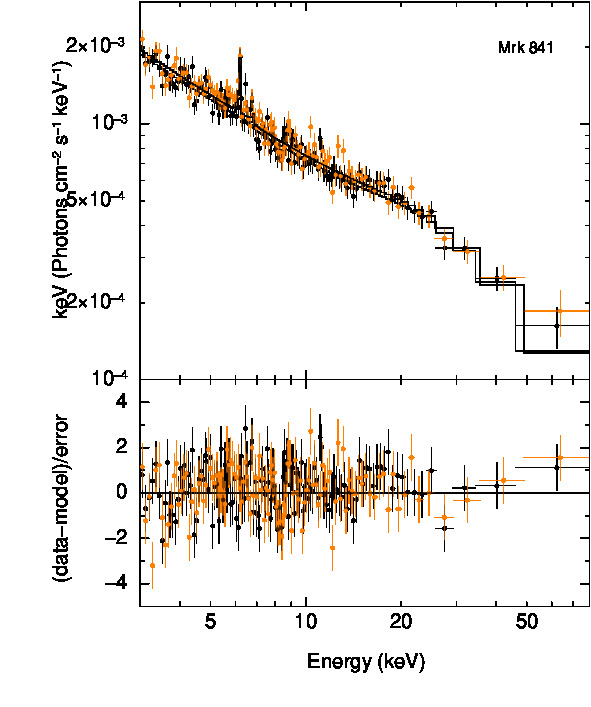}
     }
\vspace{-0.5cm}
\end{figure*}
\begin{figure*}
\vbox{
     \hspace{-0.05cm}\includegraphics[scale=0.27]{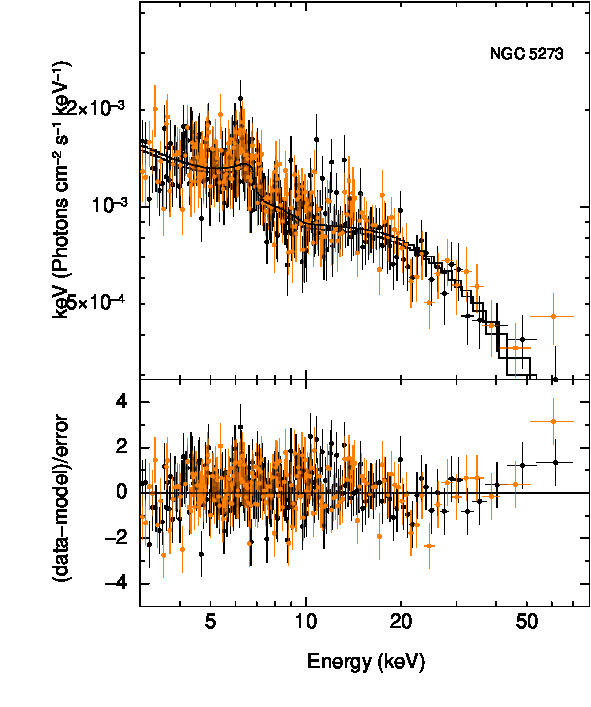}
     \hspace{-0.05cm}\includegraphics[scale=0.27]{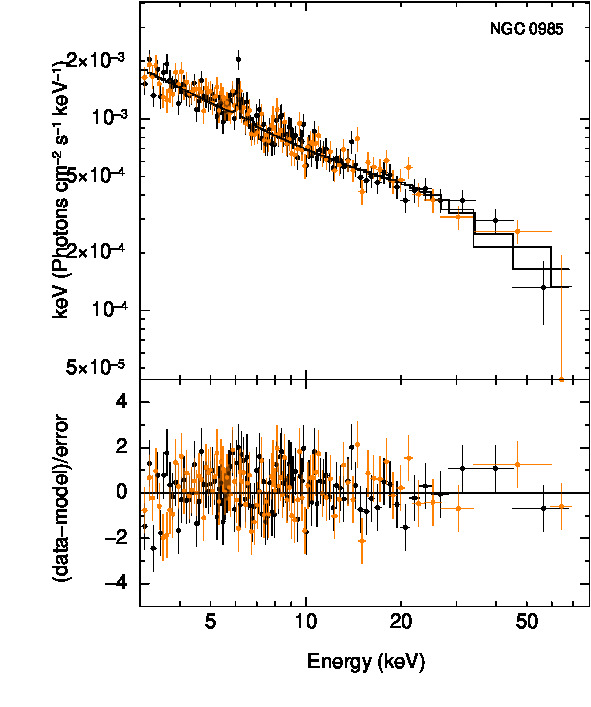}
     \hspace{-0.05cm}\includegraphics[scale=0.27]{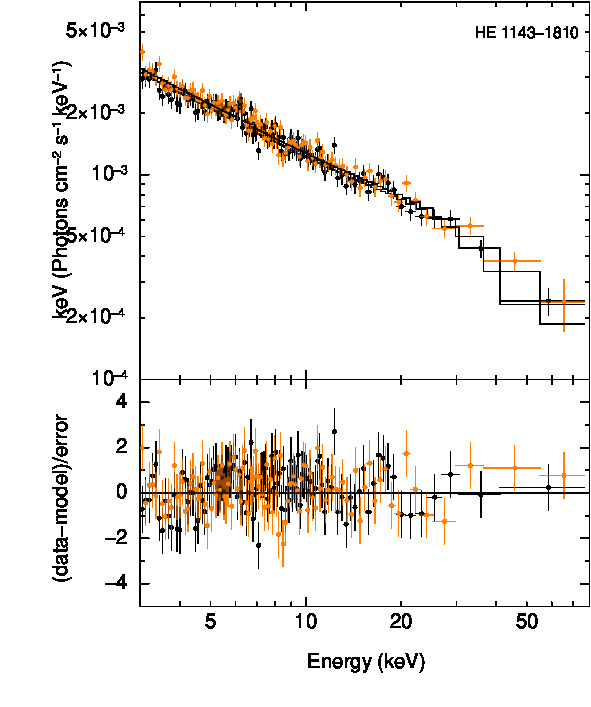}
     }
\vspace{-0.5cm}
\caption{Upper panel: The unfolded spectra and the best fit models ({\it const $\times$ TBabs $\times$ ztbabs(xillverCP/relxillCP/relxillCP+xillverCP))}) along with the residuals (lower panel) of the spectral fitting plots to the sample. Here black points are for FPMA and orange points are for FPMB. The spectra are re-binned for visualization purposes only.}
\label{figure-7}
\end{figure*}
\clearpage

\section{Other models}
\label{sec:models}
\subsection{Phenomenological model fitting}
Our first approach is a simple absorbed power-law model ({\it zpo}) fit to the data 
that has the following form in XSPEC,
\begin{equation}
const \times TBabs \times zTBabs \times (zpo+zgauss)
\end{equation}
The parameters those were treated free in the {\it zpo} model are the photon 
index ($\Gamma$) and the normalization ($N_{pow}$). For most of the 
sources, the data showed the presence of a Fe K$\alpha$ emission line
which was modelled using a {\it zgauss} component. For most 
of the observations, the line energy ($E$), the width ($\sigma$) and the normalization ($N_{line}$) were kept free during the fit. However, in few cases we noticed the necessity to freeze either $E$ or $\sigma$ or both in order to avoid $\sigma$ getting pegged at very low values during the error calculation. We note that, freezing the Fe K$\alpha$ line parameter/s significantly affected neither the fit statistics nor the derived physical parameters. Thus in few cases, we froze $E$ and $\sigma$ to 6.4 keV and 0.1 keV respectively. All the model parameters along with the fit statistic ($\chi^2/dof$) are given in Table \ref{table-2}.

Our second approach consisted of fitting the spectra with the widely used 
exponential cut-off power-law with a neutral reflection component namely 
{\it pexrav} \citep{1995MNRAS.273..837M}. Here too, we used the 
{\it zgauss} component to model the Fe K$\alpha$ emission line present in 
the sources.  In XSPEC,  the model has the form
\begin{equation}
const \times TBabs \times zTBabs \times (pexrav+zgauss)
\end{equation}
This simple model fit could provide us with constraints on important parameters
of the sources such as $\Gamma$, $\rm{E_{cut}}$ and the reflection fraction (R). 
These parameters along with the model normalization ($N_{pexrav}$) were treated 
as free parameters during the fitting. In this model, the accretion disk
is considered as a neutral one and for the neutral material we used
solar abundance. We fixed the inclination to the default of cos($\theta_{incl}$) = 0.45.
Also, the energy of the line in the {\it zgauss} component was frozen 
to the best fit values obtained from the ({\it zpo+zgauss}) model, while
$\sigma$ was treated as a free parameter in most of the cases. For
few sources, we had to freeze $\sigma$ to 0.1 keV to
better constrain the other model parameters, except for 
Mrk 915 (see Table \ref{table-3}) which showed a very narrow 
emission line. For Mrk 915, fixing $\sigma$ to 0.001 keV produced an 
acceptable fit with no significant residue at around the line emission region. All the best fit model parameters along with $\chi^2/dof$ for the fit are given 
in Table \ref{table-3}.

\subsection{Physical model fitting}
To estimate the high energy cut-off, $\rm{E_{cut}}$ we adopted the physical models {\it xillver} \citep{2010ApJ...718..695G, Garc_a_2011} and/or {\it relxill} \citep{2014ApJ...782...76G, 2014MNRAS.444L.100D}. {\it relxill} models the underlying continuum with a cut-off power law, but convolves the angle dependent {\it xillver} reflection with relativistic smearing. Here too, the spectra of 12 sources were fitted using {\it xillver}, 4 sources were modelled using {\it relxill} and in 2 sources both {\it xillver} and {\it relxill} were used. These models have the following form in XSPEC, 

\begin{enumerate}
\item const $\times$ TBabs $\times$ zTBabs $\times$ xillver
\item const $\times$ TBabs $\times$ zTBabs $\times$ relxill
\item const $\times$ TBabs $\times$ zTBabs $\times$ (relxill+xillver)
\end{enumerate}

For these models too,  we treated the parameters in the same manner as describe in Section \ref{sec:section3}. The best fit model parameters are given in Table \ref{table-4}.

\begin{table*}
\caption{Best fit parameters of the  Model {\it const $\times$ TBabs $\times$ zTBabs $\times$ (zpo+zgauss)} to the source spectra. The model normalization is in units of photons keV$^{-1}$ cm$^{-2}$s$^{-1}$, and $\rm N_{H}^{INT}$ is the host galaxy hydrogen column density in units of atoms $\rm{cm^{-2}}$. $\rm{C_{FPMA/FPMB}}$ is the cross-calibration constant. The asterisk(*) against entries indicates that they are frozen.}\label{table-2}
\centering
\begin{tabular}{p{0.11\linewidth}p{0.05\linewidth}p{0.07\linewidth}p{0.07\linewidth}p{0.07\linewidth}p{0.07\linewidth}p{0.07\linewidth}p{0.07\linewidth}p{0.07\linewidth}p{0.07\linewidth}}
\hline
Source & Epoch & $\rm N_{H}^{INT}$ & $\Gamma$ & $\rm N_{pow}$ & E & $\sigma$ & $\rm N_{line}$ & $\chi^2/dof$ & $\rm{C_{FPMA/FPMB}}$  \\
& & & & & & & & &  \\
& & ($10^{22}$)  & & (10$^{-2}$) & (keV) & (keV) & (10$^{-5}$) & &  \\
\hline
1H 0419$-$577 & A & 3.85$^{+1.10}_{-1.08}$ & 1.84$^{+0.03}_{-0.03}$ & 0.49$^{+0.04}_{-0.04}$ & - & - & - & 762/735 & 1.06$^{+0.02}_{-0.02}$ \\
& B & 3.39$^{+1.16}_{-1.14}$ & 1.83$^{+0.03}_{-0.03}$ & 0.55$^{+0.05}_{-0.04}$ & - & - & - & 671/682 & 1.02$^{+0.02}_{-0.02}$ \\
& C & 3.53$^{+0.61}_{-0.61}$ & 1.84$^{+0.02}_{-0.02}$ & 0.61$^{+0.03}_{-0.03}$ & $6.4^{*}$ & $<$0.22 & 0.53$^{+0.03}_{-0.03}$ & 1330/1156 & 1.04$^{+0.01}_{-0.01}$ \\
Mrk 915 & A & 3.09$^{+1.04}_{-1.11}$ & 1.69$^{+0.04}_{-0.04}$ & 0.24$^{+0.02}_{-0.02}$ & $6.4^{*}$ & $<$0.26 & 1.09$^{+0.35}_{-0.31}$ & 661/650 & 1.04$^{+0.02}_{-0.02}$ \\
& B & 4.64$^{+1.54}_{-1.50}$ & 1.65$^{+0.05}_{-0.05}$ & 0.15$^{+0.02}_{-0.02}$ & $6.4^{*}$ & $<$0.46 & 0.84$^{+0.78}_{-0.28}$ & 542/543 & 1.05$^{+0.03}_{-0.03}$ \\
& C & 6.45$^{+1.93}_{-1.95}$ & 1.66$^{+0.06}_{-0.06}$ & 0.13$^{+0.02}_{-0.02}$ & $6.4^{*}$ & $<$0.30 & 1.02$^{+0.40}_{-0.27}$ & 426/435 & 1.04$^{+0.03}_{-0.03}$ \\
3C 111 & A & 1.71$^{+0.93}_{-0.92}$ & 1.81$^{+0.03}_{-0.03}$ & 1.50$^{+0.12}_{-0.11}$ & 6.42$^{+0.12}_{-0.11}$ & $0.1^{*}$ & 2.81$^{+1.40}_{-1.40}$ & 662/683 & 1.01$^{+0.02}_{-0.02}$ \\
& B & 2.09$^{+0.65}_{-0.64}$ & 1.81$^{+0.02}_{-0.02}$ & 1.39$^{+0.07}_{-0.07}$ & $6.4^{*}$ & 0.20$^{+0.16}_{-0.16}$ & 2.62$^{+1.25}_{-1.13}$ & 976/949 & 1.01$^{+0.01}_{-0.01}$ \\
& C & 2.17$^{+0.73}_{-0.73}$ & 1.72$^{+0.02}_{-0.02}$ & 0.82$^{+0.05}_{-0.05}$ & 6.47$^{+0.09}_{-0.09}$ & $<$0.30 & 1.90$^{+0.99}_{-0.70}$ & 942/898 & 1.00$^{+0.01}_{-0.01}$ \\
NGC 3783 & A & - &  1.57$^{+0.01}_{-0.01}$ & 0.85$^{+0.02}_{-0.02}$ & 6.15$^{+0.06}_{-0.07}$ & 0.34$^{+0.08}_{-0.07}$ & 11.94$^{+1.79}_{-1.69}$ & 1163/1040 & 1.03$^{+0.01}_{-0.01}$ \\
& B & 1.22$^{+0.66}_{-0.67}$ & 1.52$^{+0.02}_{-0.02}$ & 0.65$^{+0.04}_{-0.04}$ & 6.22$^{+0.06}_{-0.07}$ & 0.28$^{+0.10}_{-0.09}$ & 9.95$^{+1.92}_{-1.69}$ & 1220/1004 & 1.04$^{+0.01}_{-0.01}$ \\
& C & 5.16$^{+0.99}_{-0.96}$ & 1.46$^{+0.03}_{-0.03}$ & 0.56$^{+0.05}_{-0.04}$ & 6.11$^{+0.09}_{-0.10}$ & 0.38$^{+0.12}_{-0.10}$ & 15.83$^{+3.22}_{-2.77}$ & 1068/794 & 1.04$^{+0.01}_{-0.01}$ \\
& D & 3.86$^{+0.84}_{-0.83}$ & 1.50$^{+0.03}_{-0.03}$ & 0.70$^{+0.05}_{-0.05}$ & $6.4^{*}$ & 0.16$^{+0.07}_{-0.07}$ & 9.00$^{+1.61}_{-1.49}$ & 1210/916 & 1.02$^{+0.02}_{-0.02}$ \\
NGC 7469 & A & - & 1.82$^{+0.02}_{-0.02}$ & 0.96$^{+0.04}_{-0.04}$ & 6.32$^{+0.11}_{-0.09}$ & 0.28$^{+0.30}_{-0.15}$ & 6.21$^{+2.66}_{-1.67}$ & 631/633 & 1.02$^{+0.02}_{-0.02}$  \\
& B & - & 1.76$^{+0.02}_{-0.02}$ & 0.87$^{+0.04}_{-0.04}$ & 6.39$^{+0.66}_{-0.66}$ & $<$0.1 & 3.31$^{+0.94}_{-0.92}$ & 632/629 & 1.02$^{+0.02}_{-0.02}$ \\
& C & - & 1.75$^{+0.02}_{-0.02}$ & 0.70$^{+0.03}_{-0.03}$ & 6.34$^{+0.07}_{-0.06}$ & 0.18$^{+0.10}_{-0.16}$ & 5.55$^{+1.38}_{-1.32}$ & 577/606 & 1.00$^{+0.02}_{-0.02}$ \\
& D & - & 1.79$^{+0.02}_{-0.02}$ & 0.83$^{+0.04}_{-0.03}$ & 6.35$^{+0.12}_{-0.22}$ & $<$0.46 & 2.45$^{+1.55}_{-1.07}$ & 666/634 & 1.02$^{+0.02}_{-0.02}$ \\
& E & - & 1.75$^{+0.02}_{-0.02}$ & 0.81$^{+0.04}_{-0.03}$ & 6.32$^{+0.06}_{-0.06}$ & $0.1^{*}$ & 4.22$^{+1.02}_{-1.02}$ & 682/620 & 1.01$^{+0.02}_{-0.02}$ \\
& F & - & 1.73$^{+0.02}_{-0.02}$ & 0.73$^{+0.03}_{-0.03}$ & 6.37$^{+0.12}_{-0.16}$ & 0.29$^{+0.18}_{-0.16}$ & 4.88$^{+1.84}_{-1.56}$ & 611/605 & 1.00$^{+0.02}_{-0.02}$ \\
& G & - & 1.77$^{+0.02}_{-0.02}$ & 0.93$^{+0.04}_{-0.03}$ & 6.35$^{+0.10}_{-0.10}$ & 0.25$^{+0.13}_{-0.11}$ & 5.09$^{+1.53}_{-1.38}$ & 679/695 & 1.04$^{+0.02}_{-0.02}$ \\
Mrk 110 & A & - & 1.80$^{+0.01}_{-0.01}$ & 1.18$^{+0.01}_{-0.01}$ & 6.4$^{+0.07}_{-0.07}$  & 0.34$^{+0.09}_{-0.08}$  & 3.95$^{+0.63}_{-0.68}$ & 1631/1475 & 1.01$^{+0.01}_{-0.01}$   \\
& B & - & 1.81$^{+0.01}_{-0.01}$ & 0.86$^{+0.02}_{-0.02}$ & 6.34$^{+0.11}_{-0.13}$  & 0.34$^{+0.23}_{-0.16}$  & 3.30$^{+1.12}_{-0.92}$ & 1036/1026 & 1.04$^{+0.01}_{-0.01}$   \\
& C & - & 1.76$^{+0.01}_{-0.01}$ & 0.65$^{+0.02}_{-0.02}$ & 6.35$^{+0.09}_{-0.08}$  & 0.25$^{+0.17}_{-0.15}$   & 2.74$^{+0.83}_{-0.73}$ & 1031/1007 & 1.04$^{+0.01}_{-0.01}$   \\
UGC 06728 & A & - & 1.62$^{+0.04}_{-0.04}$  & 0.15$^{+0.01}_{-0.01}$  & 6.32$^{+0.27}_{-0.26}$  & $<$0.58 & 1.31$^{+0.86}_{-0.77}$  & 189/198 & 1.03$^{+0.05}_{-0.05}$  \\
& B & - & 1.64$^{+0.02}_{-0.02}$  & 0.30$^{+0.01}_{-0.01}$  & 6.4$^{+0.09}_{-0.11}$  & $<$0.28 & 1.23$^{+0.54}_{-0.45}$  & 742/708 & 1.01$^{+0.02}_{-0.02}$  \\
NGC 4258 & A & 13.12$^{+2.38}_{-2.26}$ & 1.80$^{+0.07}_{-0.07}$ & 0.13$^{+0.03}_{-0.02}$ & - & - & - & 264/279 & 1.04$^{+0.04}_{-0.04}$ \\
& B & 14.51$^{+2.05}_{-1.96}$ & 1.89$^{+0.06}_{-0.06}$ & 0.13$^{+0.02}_{-0.02}$ & 6.4$^{+0.26}_{-0.15}$ & $0.1^{*}$ & 0.31$^{+0.19}_{-0.19}$ & 456/408 & 0.98$^{+0.03}_{-0.03}$ \\
KUG 1141+371 & A & 1.57$^{+1.28}_{-1.25}$ & 1.88$^{+0.04}_{-0.04}$ & 0.43$^{+0.05}_{-0.04}$ & - & - & - & 443/487 & 1.06$^{+0.03}_{-0.03}$ \\
& B & 2.40$^{+2.13}_{-2.04}$ & 1.76$^{+0.07}_{-0.07}$ & 0.25$^{+0.05}_{-0.04}$ & - & - & - & 241/249 & 1.05$^{+0.04}_{-0.04}$ \\
MCG-06-30-15 & A & - & 1.94$^{+0.02}_{-0.01}$ & 1.68$^{+0.05}_{-0.05}$ & $6.4^{*}$ & 0.29$^{+0.12}_{-0.09}$ & 7.43$^{+1.78}_{-1.61}$ & 785/735 & 1.02$^{+0.02}_{-0.02}$ \\
& B & - & 1.93$^{+0.01}_{-0.01}$ & 1.70$^{+0.02}_{-0.02}$ & $6.4^{*}$ & 0.28$^{+0.06}_{-0.06}$ & 6.27$^{+0.78}_{-0.75}$ & 1931/1374 & 1.03$^{+0.01}_{-0.01}$ \\
& C & - & 1.77$^{+0.02}_{-0.02}$ & 0.78$^{+0.03}_{-0.03}$ & $6.4^{*}$ & 0.26$^{+0.07}_{-0.07}$ & 6.03$^{+1.18}_{-1.13}$ & 827/743 & 1.04$^{+0.02}_{-0.02}$ \\
NGC 5506 & A & 1.11$^{+0.48}_{-0.49}$ & 1.58$^{+0.02}_{-0.02}$ & 1.07$^{+0.05}_{-0.05}$ & 6.32$^{+0.06}_{-0.06}$ & 0.45$^{+0.08}_{-0.07}$ & 18.34$^{+2.56}_{-2.27}$ & 1493/1231 & 1.00$^{+0.01}_{-0.01}$ \\
& B & 1.01$^{+0.43}_{-0.43}$ & 1.56$^{+0.01}_{-0.01}$ & 1.00$^{+0.04}_{-0.04}$ & 6.31$^{+0.03}_{-0.03}$ & 0.30$^{+0.04}_{-0.04}$ & 16.28$^{+1.52}_{-1.44}$ & 1621/1313 & 1.05$^{+0.01}_{-0.01}$ \\
& C & 0.73$^{+0.46}_{-0.49}$ & 1.59$^{+0.02}_{-0.02}$ & 1.45$^{+0.01}_{-0.01}$ & 6.22$^{+0.07}_{-0.10}$ & 0.50$^{+0.11}_{-0.09}$ & 25.97$^{+4.74}_{-3.56}$ & 1539/1296 & 1.04$^{+0.01}_{-0.01}$ \\
MCG+08-11-011 & A & - & 1.77$^{+0.01}_{-0.01}$ & 1.51$^{+0.02}_{-0.02}$ & 6.33$^{+0.04}_{-0.05}$ & 0.27$^{+0.07}_{-0.07}$ & 8.35$^{+1.17}_{-1.13}$ & 1301/1315 & 1.03$^{+0.01}_{-0.01}$ \\
& B & - & 1.70$^{+0.02}_{-0.02}$ & 1.12$^{+0.04}_{-0.04}$ & 6.36$^{+0.09}_{-0.08}$ & 0.36$^{+0.11}_{-0.11}$ & 11.97$^{+2.49}_{-2.34}$ & 753/744 & 1.01$^{+0.02}_{-0.02}$ \\
GRS 1734-292 & A & 3.64$^{+0.71}_{-0.73}$ & 1.83$^{+0.02}_{-0.02}$ & 2.54$^{+0.15}_{-0.15}$ & $6.4^{*}$ & $<$0.41 & 3.15$^{+2.44}_{-2.16}$ & 1018/865 & 1.04$^{+0.01}_{-0.01}$ \\
& B & 4.13$^{+0.74}_{-0.78}$ & 1.76$^{+0.02}_{-0.02}$ & 1.72$^{+0.10}_{-0.10}$ & 6.07$^{+0.15}_{-0.16}$ & $<$0.36 & 3.48$^{+2.34}_{-1.70}$ & 869/855 & 0.98$^{+0.01}_{-0.01}$ \\
Mrk 926 & A & - & 1.76$^{+0.01}_{-0.01}$ & 1.59$^{+0.02}_{-0.02}$ & 6.32$^{+0.08}_{-0.08}$ & 0.38$^{+0.10}_{-0.09}$ & 6.26$^{+1.23}_{-1.14}$ & 1453/1467 & 1.00$^{+0.01}_{-0.01}$ \\
& B & - & 1.69$^{+0.03}_{-0.03}$ & 0.50$^{+0.03}_{-0.03}$ & 6.29$^{+0.12}_{-0.13}$ & $<$0.29 & 2.55$^{+1.24}_{-1.18}$ & 465/493 & 1.06$^{+0.03}_{-0.03}$ \\
Mrk 841 & A  & - & 1.75$^{+0.03}_{-0.03}$ & 0.48$^{+0.03}_{-0.04}$ & 6.33$^{+0.28}_{-0.85}$ & 0.44$^{+0.73}_{-0.22}$ & 2.98$^{+4.39}_{-1.41}$ & 477/506 & 1.00$^{+0.03}_{-0.03}$ \\
& B & - & 1.73$^{+0.02}_{-0.02}$ & 0.42$^{+0.02}_{-0.02}$ & 6.33$^{+0.16}_{-0.40}$ & 0.41$^{+0.43}_{-0.17}$ & 2.93$^{+2.16}_{-0.94}$ & 750/743 & 1.04$^{+0.02}_{-0.02}$ \\
NGC 5273 & A & - & 1.54$^{+0.02}_{-0.02}$ & 0.32$^{+0.02}_{-0.02}$ & 6.4* & 0.32$^{+0.22}_{-0.15}$ & 4.25$^{+1.58}_{-1.32}$ & 634/576 & 1.07$^{+0.03}_{-0.03}$ \\
& B & - & 1.42$^{+0.03}_{-0.03}$ & 0.24$^{+0.01}_{-0.01}$ & 6.25$^{+0.09}_{-0.09}$ & 0.24$^{+0.25}_{-0.14}$ & 4.92$^{+1.91}_{-1.36}$ & 483/507 & 1.03$^{+0.03}_{-0.03}$ \\
NGC 0985 & A & 5.22$^{+2.73}_{-2.70}$ & 1.70$^{+0.08}_{-0.08}$ & 0.29$^{+0.07}_{-0.06}$ & 6.4* & 0.31$^{+0.49}_{-0.27}$ & 1.99$^{+1.99}_{-1.27}$ & 308/266 & 1.08$^{+0.05}_{-0.04}$ \\
& B & $<$1.68 & 1.74$^{+0.03}_{-0.03}$ & 0.42$^{+0.03}_{-0.03}$ & 6.33$^{+0.15}_{-0.19}$ & 0.24$^{+0.19}_{-0.13}$ & 2.79$^{+1.22}_{-1.06}$ & 388/465 & 1.00$^{+0.03}_{-0.03}$ \\
HE 1143$-$1810 & A & 1.91$^{+1.39}_{-1.37}$ & 1.82$^{+0.04}_{-0.04}$ & 0.71$^{+0.08}_{-0.08}$ & 6.4* & $<$0.99 & 1.67$^{+1.83}_{-1.59}$ & 500/506 & 1.03$^{+0.03}_{-0.02}$ \\
& B & $<$2.81 & 1.76$^{+0.05}_{-0.05}$ & 0.60$^{+0.08}_{-0.07}$ & 6.36$^{+0.15}_{-0.16}$ & 0.26$^{+0.20}_{-0.16}$ & 3.43$^{+1.71}_{-1.41}$ & 466/485 & 1.01$^{+0.03}_{-0.03}$ \\
& C & - & 1.77$^{+0.02}_{-0.02}$ & 0.76$^{+0.03}_{-0.03}$ & 6.24$^{+0.17}_{-0.21}$ & 3.75$^{+2.15}_{-1.55}$ & 1.04$^{+0.02}_{-0.02}$ & 530/594 & 1.04$^{+0.02}_{-0.02}$ \\
& D & 1.54$^{+1.14}_{-1.11}$ & 1.82$^{+0.04}_{-0.04}$ & 0.85$^{+0.09}_{-0.08}$ & 6.4$^{+0.17}_{-0.15}$ & 0.10* & 1.74$^{+1.01}_{-1.01}$ & 594/560 & 1.04$^{+0.02}_{-0.02}$ \\
& E & 1.82$^{+1.11}_{-1.08}$ & 1.83$^{+0.04}_{-0.04}$ & 0.86$^{+0.08}_{-0.08}$ & 6.4* & 0.10* & 1.05$^{+0.02}_{-0.02}$ & 597/569 & 1.05$^{+0.02}_{-0.02}$ \\
\hline\hline
\end{tabular}
\end{table*}

\begin{table*}
\caption{Best fit parameters of the  Model {\it const $\times$ TBabs $\times$ zTBabs $\times$ (pexrav+zgauss)} to the source spectra. The model normalization is in units of keV$^{-1}$ cm$^{-2}$s$^{-1}$, and $\rm N_{H}^{INT}$ is the host galaxy hydrogen column density in units of atoms $\rm{cm^{-2}}$. The asterisk(*) against entries indicates that they are frozen.}\label{table-3}
\centering
\begin{tabular}{p{0.11\linewidth}p{0.05\linewidth}p{0.07\linewidth}p{0.07\linewidth}p{0.07\linewidth}p{0.07\linewidth}p{0.07\linewidth}p{0.07\linewidth}p{0.07\linewidth}p{0.07\linewidth}p{0.07\linewidth}}
\hline
Source & Epoch & $\rm N_{H}^{INT}$ & $\Gamma$  & $\rm{E_{cut}}$ & R & $\rm N_{pexrav}$ & E & $\sigma$ & $\rm N_{line}$ & $\chi^2/dof$   \\
& & & & & & & & &  \\
& & ($10^{22}$)  & &  (keV) &  & (10$^{-2}$) & (keV) & (keV) & (10$^{-5}$) &   \\
\hline
1H 0419$-$577 & A & 2.90$^{+1.54}_{-1.54}$ & 1.82$^{+0.11}_{-0.11}$ & 94$^{+82}_{-31}$ & 0.46$^{+0.30}_{-0.25}$ & 0.38$^{+0.08}_{-0.06}$ & - & - & - & 739/733  \\
& B & 1.96$^{+1.66}_{-1.65}$ & 1.74$^{+0.12}_{-0.12}$ & 88$^{+82}_{-30}$ & 0.28$^{+0.29}_{-0.24}$ & 0.39$^{+0.08}_{-0.07}$ & - & - & - & 657/680  \\
& C & $<$1.05 & 1.62$^{+0.07}_{-0.02}$ & 51$^{+9}_{-4}$ & 0.33$^{+0.15}_{-0.11}$ & 0.34$^{+0.05}_{-0.01}$ & $6.4^{*}$ & $<$0.34 & 1.05$^{+0.39}_{-0.43}$ & 1167/1154  \\
Mrk 915 & A & 4.66$^{+1.29}_{-1.30}$ & 1.87$^{+0.10}_{-0.10}$ & $>$667 & 0.53$^{+0.35}_{-0.30}$ & 0.31$^{+0.06}_{-0.05}$ & $6.4^{*}$ & $0.001^{*}$ & 0.92$^{+0.32}_{-0.32}$ & 650/649 \\
& B & 3.76$^{+2.28}_{-2.25}$ & 1.58$^{+0.17}_{-0.17}$ & 112$^{+862}_{-55}$ & $<$0.62 & 0.13$^{+0.05}_{-0.03}$ & $6.4^{*}$ & $0.001^{*}$ & 0.84$^{+0.28}_{-0.29}$ & 539/542 \\
& C & 6.33$^{+2.76}_{-2.73}$ & 1.71$^{+0.22}_{-0.22}$ & 84$^{+351}_{-40}$ & 0.74$^{+0.81}_{-0.53}$ & 0.13$^{+0.06}_{-0.04}$ & $6.4^{*}$ & $0.001^{*}$ & 0.93$^{+0.29}_{-0.28}$ & 419/434 \\
3C 111 & A & 1.74$^{+0.92}_{-0.69}$ & 1.81$^{+0.04}_{-0.08}$ & $>$864 & $<$0.11 & 1.38$^{+0.18}_{-0.18}$ & $6.42^{*}$ & $<$0.20 & 2.67$^{+1.31}_{-1.27}$ & 661/681  \\
& B & 0.73$^{+0.81}_{-0.70}$ & 1.68$^{+0.05}_{-0.05}$ & 128$^{+51}_{-37}$ & $<$0.06 & 1.04$^{+0.10}_{-0.11}$ & $6.30^{*}$ & $0.1^{*}$ & 2.60$^{+0.08}_{-0.08}$ & 957/948 \\
& C & 1.33$^{+1.07}_{-1.12}$ & 1.65$^{+0.06}_{-0.06}$ & 251$^{+513}_{-130}$ & $<$0.04 & 0.68$^{+0.08}_{-0.08}$ & $6.46^{*}$ & $<$0.37 & 2.17$^{+1.14}_{-0.08}$ & 938/897  \\
NGC 3783 & A & - & 1.72$^{+0.04}_{-0.04}$ & 245$^{+150}_{-70}$ & 0.86$^{+0.21}_{-0.18}$ & 1.02$^{+0.06}_{-0.06}$ & $6.15^{*}$ & 0.33$^{+0.07}_{-0.06}$ & 11.36$^{+1.62}_{-1.54}$ & 1049/1039 \\
& B & 3.66$^{+0.67}_{-1.07}$ & 1.92$^{+0.06}_{-0.11}$ & $>$284 & 1.83$^{+0.41}_{-0.42}$ & 1.17$^{+0.01}_{-0.02}$ & $6.21^{*}$ & 0.20$^{+0.07}_{-0.07}$ & 6.87$^{+1.53}_{-1.37}$ & 1030/1003 \\
& C & 7.91$^{+0.93}_{-1.82}$ & 1.91$^{+0.04}_{-0.17}$ & $>$242 & 2.00$^{+0.53}_{-0.60}$ & 1.08$^{+0.01}_{-0.03}$ & $6.11^{*}$ & 0.31$^{+0.08}_{-0.06}$ & 11.77$^{+2.88}_{-1.99}$ & 938/793 \\
& D & 6.85$^{+0.82}_{-0.81}$ & 1.94$^{+0.06}_{-0.06}$ & $>$1225 & 1.91$^{+0.47}_{-0.42}$ & 1.37$^{+0.01}_{-0.01}$ & $6.4^{*}$ & $<$0.15 & 6.21$^{+1.40}_{-1.24}$ & 1071/914 \\
NGC 7469 & A & - & 1.94$^{+0.07}_{-0.06}$ & 270$^{+2404}_{-134}$ & 0.66$^{+0.35}_{-0.28}$ & 1.08$^{+0.10}_{-0.09}$ & $6.32^{*}$ & $0.1^{*}$ & 4.48$^{+0.98}_{-0.98}$ & 614/633 \\
& B & - & 1.88$^{+0.02}_{-0.06}$ & $>$343 & 0.46$^{+0.30}_{-0.09}$ & 0.98$^{+0.07}_{-0.08}$ & $6.39^{*}$ & $0.1^{*}$ & 3.36$^{+0.10}_{-0.10}$ & 622/629 \\
& C & - & 1.93$^{+0.07}_{-0.07}$ & $>$255 & 0.77$^{+0.41}_{-0.31}$ & 0.86$^{+0.08}_{-0.08}$ & $6.34^{*}$ & $0.1^{*}$ & 4.78$^{+0.93}_{-0.93}$ & 533/606 \\
& D & - & 1.98$^{+0.05}_{-0.04}$ & $>$890 & 0.81$^{+0.34}_{-0.28}$ & 1.02$^{+0.05}_{-0.05}$ & $6.36^{*}$ & $0.1^{*}$ & 2.23$^{+0.83}_{-0.92}$ & 630/635 \\
& E & - & 1.89$^{+0.06}_{-0.06}$ & $>$523 & 0.57$^{+0.33}_{-0.26}$ & 0.95$^{+0.08}_{-0.08}$ & $6.33^{*}$ & $0.1^{*}$ & 4.12$^{+1.01}_{-1.02}$ & 663/619 \\
& F & - & 1.86$^{+0.07}_{-0.06}$ & $>$867 & 0.53$^{+0.32}_{-0.24}$ & 0.86$^{+0.08}_{-0.07}$ & $6.37^{*}$ & $0.1^{*}$ & 3.50$^{+0.97}_{-0.97}$ & 597/605 \\
& G & - & 1.87$^{+0.02}_{-0.05}$ & $>$425 & 0.38$^{+0.24}_{-0.12}$ & 1.04$^{+0.01}_{-0.07}$ & $6.35^{*}$ & $0.1^{*}$ & 3.80$^{+0.94}_{-0.95}$ & 671/695 \\
Mrk 110 & A & - & 1.72$^{+0.01}_{-0.01}$ & 132$^{+27}_{-17}$ & $<$0.03 & 0.99$^{+0.02}_{-0.01}$ & $6.4^{*}$ & 0.30$^{+0.08}_{-0.07}$ & 3.43$^{+0.31}_{-0.58}$ & 1520/1474 \\
& B & - & 1.77$^{+0.03}_{-0.03}$ & 173$^{+125}_{-51}$ & $<$0.18 & 0.77$^{+0.03}_{-0.03}$ & $6.33^{*}$ & $<$0.38 & 3.12$^{+0.67}_{-0.67}$ & 1026/1026  \\
& C & - & 1.72$^{+0.03}_{-0.03}$ & 213$^{+251}_{-75}$ & $<$0.17 & 0.59$^{+0.03}_{-0.03}$ & $6.35^{*}$ & 0.21$^{+0.17}_{-0.15}$ & 2.48$^{+0.83}_{-0.66}$ & 1021/1006  \\
UGC 06728 & A & - & 1.62$^{+0.17}_{-0.15}$ & 106$^{+8766}_{-55}$ & $<$1.28 & 0.15$^{+0.04}_{-0.03}$ & $6.32^{*}$ & $<$0.61 & 1.21$^{+0.09}_{-0.08}$ & 186/197 \\
& B & - & 1.76$^{+0.06}_{-0.06}$ & 295$^{+2468}_{-146}$ & 0.64$^{+0.32}_{-0.26}$ & 0.65$^{+0.03}_{-0.03}$ & $6.4^{*}$ & $0.1^{*}$ & 1.15$^{+0.40}_{-0.40}$ & 719/708 \\
NGC 4258 & A & 11.28$^{+3.21}_{-3.14}$ & 1.68$^{+0.25}_{-0.25}$ & 52$^{+123}_{-23}$ & $<$1.36 & 0.11$^{+0.05}_{-0.04}$ & - & - & - & 258/277 \\
& B & 10.71$^{+2.93}_{-2.89}$ & 1.57$^{+0.24}_{-0.23}$ & 32$^{+23}_{-9}$ & $<$1.10 & 0.08$^{+0.04}_{-0.03}$ & $6.4^{*}$ & $<$0.35 & 0.40$^{+0.25}_{-0.24}$  & 439/406 \\
KUG 1141+371 & A & 2.67$^{+1.59}_{-1.74}$ & 2.06$^{+0.13}_{-0.16}$ & $>$98 & 0.75$^{+0.663}_{-0.46}$ & 0.52$^{+0.12}_{-0.12}$ & - & - & - & 434/485 \\
& B & 3.76$^{+1.82}_{-2.91}$ & 1.93$^{+0.08}_{-0.23}$ & $>$86 & $<$1.44 & 0.30$^{+0.10}_{-0.10}$ & - & - & - & 238/247 \\
MCG-06-30-15 & A & - & 2.25$^{+0.09}_{-0.07}$ & $>$210 & 1.88$^{+0.67}_{-0.49}$ & 2.38$^{+0.02}_{-0.02}$ & $6.4^{*}$ & $0.1^{*}$ & 4.44$^{+1.08}_{-1.08}$ & 686/734 \\
& B & - & 2.15$^{+0.04}_{-0.03}$ & $>$405 & 1.17$^{+0.27}_{-0.15}$ & 2.18$^{+0.06}_{-0.07}$ & $6.4^{*}$ & 0.37$^{+0.40}_{-0.13}$ & 6.53$^{+3.03}_{-1.23}$ & 1541/1372 \\
& C & - & 2.05$^{+0.08}_{-0.07}$ & $>$273 & 1.49$^{+0.51}_{-0.39}$ & 1.16$^{+0.10}_{-0.09}$ & $6.4^{*}$ & 0.27$^{+0.09}_{-0.08}$ & 5.61$^{+1.26}_{-1.18}$ & 733/741 \\
NGC 5506 & A & 3.11$^{+0.50}_{-0.52}$ & 1.85$^{+0.04}_{-0.04}$ & $>$1389 & 0.97$^{+0.17}_{-0.16}$ & 1.64$^{+0.11}_{-0.10}$ & $6.32^{*}$ & 0.46$^{+0.10}_{-0.10}$ & 15.50$^{+2.57}_{-2.35}$ & 1303/1230 \\
& B & 3.42$^{+0.44}_{-0.44}$ & 1.90$^{+0.03}_{-0.03}$ & $>$2818 & 1.30$^{+0.19}_{-0.18}$ & 1.68$^{+0.09}_{-0.08}$ & $6.31^{*}$ & 0.22$^{+0.05}_{-0.05}$ & 12.23$^{+1.42}_{-1.35}$ & 1302/1312 \\
& C & 3.01$^{+0.46}_{-0.49}$ & 1.90$^{+0.03}_{-0.04}$ & $>$1364 & 1.12$^{+0.18}_{-0.16}$ & 2.35$^{+0.14}_{-0.15}$ & $6.22^{*}$ & 0.48$^{+0.09}_{-0.08}$ & 20.37$^{+3.11}_{-2.88}$ & 1284/1295 \\
MCG+08-11-011 & A & - & 1.85$^{+0.02}_{-0.02}$ & 203$^{+66}_{-41}$ & 0.36$^{+0.05}_{-0.05}$ & 1.64$^{+0.05}_{-0.04}$ & $6.33^{*}$ & 0.27$^{+0.07}_{-0.07}$ & 8.14$^{+1.18}_{-1.12}$ & 1271/1314 \\
& B & - & 1.78$^{+0.05}_{-0.05}$ & 110$^{+48}_{-27}$ & 0.52$^{+0.12}_{-0.11}$ & 1.24$^{+0.08}_{-0.08}$ & $6.28^{*}$ & 0.36$^{+0.11}_{-0.11}$ & 11.62$^{+2.40}_{-2.28}$ & 742/743 \\
GRS 1734-292 & A & 1.80$^{+1.15}_{-1.20}$ & 1.74$^{+0.10}_{-0.10}$ & 68$^{+23}_{-15}$ & 0.53$^{+0.25}_{-0.22}$ & 2.02$^{+0.36}_{-0.32}$ & $6.4^{*}$ & 0.24$^{+0.22}_{-0.17}$ & 4.17$^{+2.98}_{-2.45}$ & 938/863 \\
& B & 3.50$^{+1.16}_{-1.27}$ & 1.79$^{+0.10}_{-0.10}$ & 108$^{+64}_{-31}$ & 0.65$^{+0.26}_{-0.23}$ & 1.67$^{+0.30}_{-0.28}$ & $6.06^{*}$ & $<$0.35 & 3.53$^{+2.53}_{-1.92}$ & 805/854 \\
Mrk 926 & A & - & 1.75$^{+0.02}_{-0.02}$ & 179$^{+52}_{-33}$ & 0.12$^{+0.03}_{-0.03}$ & 1.52$^{+0.04}_{-0.04}$ & $6.32^{*}$ & 0.36$^{+0.10}_{-0.09}$ & 5.92$^{+1.18}_{-1.11}$ & 1433/1466 \\
& B & - & 1.70$^{+0.08}_{-0.08}$ & 154$^{+1367}_{-77}$ & 0.20$^{+0.18}_{-0.15}$ & 0.50$^{+0.06}_{-0.05}$ & $6.29^{*}$ & $<$0.29 & 2.57$^{+1.24}_{-1.12}$ & 464/493 \\
Mrk 841 & A & - & 1.83$^{+0.08}_{-0.08}$ & 185$^{+1658}_{-93}$ & 0.34$^{+0.18}_{-0.16}$ & 0.51$^{+0.06}_{-0.05}$ & $6.33^{*}$ & 0.48$^{+0.59}_{-0.26}$ & 2.98$^{+2.19}_{-1.49}$ & 470/505 \\
& B & - & 1.84$^{+0.05}_{-0.05}$ & 173$^{+213}_{-65}$ & 0.47$^{+0.13}_{-0.12}$  & 0.47$^{+0.03}_{-0.03}$ & $6.33^{*}$ & 0.54$^{+0.42}_{-0.28}$  & 3.20$^{+1.70}_{-1.21}$ & 727/742 \\
NGC 5273 & A & - & 1.68$^{+0.07}_{-0.07}$ & 182$^{+332}_{-77}$ & 0.55$^{+0.19}_{-0.17}$ & 0.39$^{+0.04}_{-0.04}$ & $6.4^{*}$ & 0.42$^{+0.41}_{-0.24}$ & 4.61$^{+2.46}_{-1.70}$ & 618/574 \\
& B & - & 1.44$^{+0.08}_{-0.08}$ & 59$^{+24}_{-14}$ & 0.61$^{+0.21}_{-0.18}$ & 0.25$^{+0.03}_{-0.03}$ & $6.25^{*}$ & 0.21$^{+0.25}_{-0.13}$ & 4.59$^{+1.82}_{-1.26}$ & 462/506 \\
NGC 0985 & A & 8.61$^{+2.36}_{-3.07}$ & 2.15$^{+0.55}_{-0.38}$ & $>$176 & 1.89$^{+0.39}_{-0.79}$ & 0.53$^{+0.12}_{-0.11}$ & $6.4^{*}$ & $0.20^{*}$ & 0.89$^{+0.93}_{-0.89}$ & 296/266 \\
& B & - & 1.83$^{+0.10}_{-0.10}$ & $>$160 & 0.37$^{+0.47}_{-0.30}$ & 0.44$^{+0.05}_{-0.06}$ & $6.33^{*}$ & 0.26$^{+0.22}_{-0.12}$ & 2.81$^{+1.29}_{-0.99}$ & 383/463 \\
HE 1134$-$1810 & A & $<$3.44 & 1.80$^{+0.16}_{-0.16}$ & 95$^{+174}_{-39}$ & 0.43$^{+0.47}_{-0.34}$ & 0.65$^{+0.21}_{-0.15}$ & $6.4^{*}$ & $0.10^{*}$ & $<$1.66 & 491/505 \\
& B & $<$3.37 & 1.76$^{+0.18}_{-0.13}$ & $>$82 & $<$0.68 & 0.56$^{+0.21}_{-0.11}$ & $6.36^{*}$ & 0.26$^{+0.20}_{-0.17}$ & 3.40$^{+1.81}_{-1.56}$ & 462/484 \\
& C & - & 1.76$^{+0.07}_{-0.07}$ & 154$^{+239}_{-60}$ & 0.25$^{+0.27}_{-0.22}$ & 0.72$^{+0.07}_{-0.06}$ & $6.24^{*}$ & 0.28$^{+0.35}_{-0.22}$ & 3.44$^{+2.04}_{-1.48}$ & 522/595 \\
& D & $<$3.37 & 1.85$^{+0.14}_{-0.14}$ & $>$103 & 0.27$^{+0.36}_{-0.27}$ & 0.85$^{+0.23}_{-0.18}$ & $6.4^{*}$ & $0.10^{*}$ & 1.61$^{+1.05}_{-1.07}$ & 590/559 \\
& E & 2.27$^{+1.64}_{-1.65}$ & 1.92$^{+0.14}_{-0.14}$ & $>$102 & 0.49$^{+0.42}_{-0.32}$ & 0.92$^{+0.24}_{-0.19}$ & $6.4^{*}$ & $0.10^{*}$ & $<$1.33 & 588/567 \\
\hline\hline
\end{tabular}
\end{table*}

\begin{table*}
\caption{Best fit parameters of the  Model {\it const $\times$ TBabs $\times$ zTBabs $\times$ (xillver/relxill/(relxill+xillver))} to the source spectra. The model normalization is in units of photons keV$^{-1}$ cm$^{-2}$s$^{-1}$, and $\rm N_{H}^{INT}$ is the host galaxy hydrogen column density in units of atoms $\rm{cm^{-2}}$. The asterisk(*) against entries indicates that they are frozen.}\label{table-4}
\centering
\begin{tabular}{p{0.11\linewidth}p{0.03\linewidth}p{0.06\linewidth}p{0.06\linewidth}p{0.06\linewidth}p{0.06\linewidth}p{0.06\linewidth}p{0.06\linewidth}p{0.06\linewidth}p{0.06\linewidth}p{0.06\linewidth}p{0.06\linewidth}}
\hline
Source & Epoch & $\rm N_{H}^{INT}$ & $\Gamma$ & $\rm{E_{cut}}$ & R & $log\xi$ & $\rm AF_{e}$ & $\beta_{1}$ & $\rm N_{xillver}$ & $\rm N_{relxill}$ & $\chi^2/dof$   \\
& & & & & & & & & & & \\
& & ($10^{22}$) & & (keV) & & & & &  ($10^{-4}$) & ($10^{-4}$) & \\
\hline
1H 0419$-$577 & A & 1.59$^{+0.67}_{-0.66}$ & 1.70$^{+0.02}_{-0.03}$ & 72$^{+9}_{-8}$ & 0.20$^{+0.11}_{-0.10}$ & - & - & - & 0.64$^{+0.10}_{-0.13}$ & - &  739/733 \\
& B & 1.04$^{+0.72}_{-0.70}$ & 1.66$^{+0.02}_{-0.03}$ & 72$^{+15}_{-8}$ & 0.13$^{+0.11}_{-0.12}$ & - & - & - & 0.74$^{+0.58}_{-0.16}$ & - &  657/680 \\
& C & $<$0.38 & 1.59$^{+0.02}_{-0.01}$ & 47$^{+2}_{-2}$ & 0.25$^{+0.06}_{-0.06}$ & - & - & - & 0.70$^{+0.06}_{-0.15}$ & - & 1173/1156 \\
Mrk 915 & A & 5.21$^{+0.74}_{-0.74}$ & 1.88$^{+0.04}_{-0.03}$ & $>$437 & 0.46$^{+0.17}_{-0.14}$ & - & - & - & 0.75$^{+0.10}_{-0.02}$ & - & 653/650 \\
& B & 4.60$^{+0.93}_{-0.90}$ & 1.65$^{+0.03}_{-0.04}$ & 70$^{+15}_{-10}$ & 0.51$^{+0.22}_{-0.21}$ & - & - & - & 0.31$^{+0.01}_{-0.01}$ & - & 544/543 \\
& C & 5.84$^{+1.13}_{-1.09}$ & 1.67$^{+0.03}_{-0.04}$ & 45$^{+5}_{-4}$ & 1.01$^{+0.39}_{-0.35}$ & - & - & - & 0.21$^{+0.07}_{-0.07}$ & - & 420/435 \\
3C 111 & A & 2.50$^{+0.58}_{-0.57}$ & 1.86$^{+0.02}_{-0.02}$ & $>$386 & 0.11$^{+0.10}_{-0.09}$ & - & - & - & 3.81$^{+0.11}_{-0.07}$ & - &  668/683 \\
& B & 1.67$^{+0.38}_{-0.39}$ & 1.75$^{+0.01}_{-0.01}$ & 133$^{+26}_{-20}$ & 0.11$^{+0.07}_{-0.06}$ & - & - & - & 2.48$^{+0.03}_{-0.01}$ & - &  974/949 \\
& C & 2.45$^{+0.45}_{-0.44}$ & 1.72$^{+0.02}_{-0.01}$ & 300$^{+170}_{-77}$ & $<$0.15 & - & - & - & 2.27$^{+0.03}_{-0.03}$ & - & 959/899 \\
NGC 3783 & A & - & 1.76$^{+0.06}_{-0.10}$ & 186$^{+104}_{-63}$ & 0.78$^{+0.35}_{-0.33}$ & 2.75$^{+0.31}_{-0.09}$ & $1.00^{*}$ & 2.74$^{+0.31}_{-0.34}$ & 0.95$^{+0.41}_{-0.45}$ & 2.05$^{+0.22}_{-0.28}$ & 1097/1038 \\
& B & 3.72$^{+0.93}_{-0.72}$ & 1.95$^{+0.03}_{-0.10}$ & $>$363 & 1.41$^{+0.56}_{-0.43}$ & 2.74$^{+0.28}_{-0.17}$   & $1.00^{*}$ & 3.45$^{+0.89}_{-0.45}$ & 1.90$^{+0.50}_{-0.54}$ & 1.83$^{+0.23}_{-0.24}$ & 1033/1002 \\
& C & 8.62$^{+1.49}_{-1.12}$ & 1.96$^{+0.05}_{-0.07}$ & $>$436 & 1.49$^{+0.85}_{-0.60}$ & 2.77$^{+0.31}_{-0.40}$ & $1.00^{*}$ & 3.22$^{+0.85}_{-0.48}$ & 2.41$^{+0.63}_{-0.71}$ & 1.69$^{+0.33}_{-0.34}$ & 937/792 \\
& D & 6.64$^{+1.17}_{-0.99}$ & 1.94$^{+0.05}_{-0.07}$ & $>$441 & 1.66$^{+0.84}_{-0.59}$ & 2.79$^{+0.25}_{-0.30}$ & $1.00^{*}$ & 3.37$^{+0.68}_{-0.45}$ & 2.20$^{+0.59}_{-0.70}$ &  1.91$^{+0.37}_{-0.36}$ & 1048/913 \\
NGC 7469 & A & - & 1.94$^{+0.02}_{-0.02}$ & 119$^{+26}_{-20}$ & 0.77$^{+0.20}_{-0.17}$ & - & - & - & 1.57$^{+0.03}_{-0.03}$ & - & 612/634 \\
& B & - & 1.88$^{+0.02}_{-0.02}$ & 301$^{+239}_{-101}$ & 0.52$^{+0.17}_{-0.16}$ & - & - & - & 1.87$^{+0.03}_{-0.04}$ & - & 622/630 \\
& C & - & 1.93$^{+0.02}_{-0.02}$ & 163$^{+56}_{-33}$ & 0.93$^{+0.20}_{-0.17}$ & - & - & - & 1.35$^{+0.03}_{-0.03}$ & - & 562/607 \\
& D & - & 1.94$^{+0.02}_{-0.02}$ & $>$463 & 0.54$^{+0.14}_{-0.14}$ & - & - & - & 1.94$^{+0.05}_{-0.03}$ & - & 625/635 \\
& E & - & 1.90$^{+0.02}_{-0.02}$ & 365$^{+402}_{-130}$ & 0.59$^{+0.19}_{-0.16}$ & - & - & - & 1.84$^{+0.03}_{-0.03}$ & - & 672/620 \\
& F & - & 1.89$^{+0.02}_{-0.02}$ & $>$407 & 0.56$^{+0.17}_{-0.16}$ & - & - & - & 2.00$^{+0.07}_{-0.04}$ & - & 599/606 \\
& G & - & 1.88$^{+0.02}_{-0.02}$ & 300$^{+200}_{-88}$ & 0.46$^{+0.14}_{-0.13}$ & - & - & - & 2.04$^{+0.03}_{-0.02}$ & - & 679/696 \\
Mrk 110 & A & - & 1.69$^{+0.02}_{-0.02}$ & 94$^{+13}_{-10}$ & 0.13$^{+0.02}_{-0.02}$ & 3.00$^{+0.04}_{-0.03}$ & $>$8.76 & $3.0^{*}$ & - & 2.02$^{+0.07}_{-0.06}$ & 1590/1474 \\
& B & - & 1.72$^{+0.03}_{-0.04}$ & 98$^{+30}_{-18}$ & 0.24$^{+0.10}_{-0.09}$ & 2.91$^{+0.32}_{-0.29}$ & $>$6.44 & $3.0^{*}$ & - & 1.43$^{+0.10}_{-0.07}$ & 1027/1025 \\
& C & - & 1.67$^{+0.02}_{-0.04}$ & 112$^{+33}_{-24}$ & 0.22$^{+0.10}_{-0.07}$ & 3.00$^{+0.24}_{-0.29}$ & $>$6.92 & $3.0^{*}$ & - & 1.27$^{+0.09}_{-0.08}$ & 1046/1006 \\
UGC 06728 & A & - & 1.61$^{+0.12}_{-0.06}$ & 91$^{+164}_{-43}$ & 0.36$^{+0.51}_{-0.21}$ & $<$3.07  & $1.00^{*}$ & $<$2.58 & - & 0.34$^{+0.11}_{-0.07}$ & 188/197 \\
& B & - & 1.65$^{+0.10}_{-0.07}$ &  252$^{+400}_{-88}$ & 0.54$^{+0.34}_{-0.22}$ & $<$3.52  & $1.00^{*}$ & $<$2.69 & - & 0.75$^{+0.26}_{-0.32}$ & 718/707 \\
NGC 4258 & A & 9.73$^{+1.43}_{-1.37}$ & 1.56$^{+0.05}_{-0.04}$ & 41$^{+5}_{-6}$ & $<$0.65 & - & - & - & 0.17$^{+0.06}_{-0.05}$ & - & 255/277 \\
& B & 11.17$^{+1.18}_{-1.14}$ & 1.60$^{+0.10}_{-0.04}$ & 30$^{+3}_{-3}$ & $<$0.83 & - & - & - & 0.13$^{+0.05}_{-0.05}$ & - & 439/408 \\
KUG 1141+371 & A & 1.23$^{+0.80}_{-0.78}$ & 1.91$^{+0.03}_{-0.04}$ & 133$^{+64}_{-34}$ & 0.39$^{+0.21}_{-0.19}$ & - & - & - & 0.63$^{+0.10}_{-0.07}$ & - & 430/485 \\
& B & 2.69$^{+1.33}_{-1.29}$ & 1.81$^{+0.05}_{-0.05}$ & $>$171 & $<$0.45 & - & - & - & 0.66$^{+0.24}_{-0.28}$ & - & 239/247 \\
MCG-06-30-15 & A & - & 1.99$^{+0.13}_{-0.09}$ & $>$183 & 1.76$^{+2.58}_{-0.49}$ & 3.09$^{+0.17}_{-0.18}$ & 1.01$^{+1.22}_{-0.34}$ & 2.83$^{+0.28}_{-0.15}$ & - & 1.46$^{+0.53}_{-0.73}$ & 647/732 \\
& B & - & 2.06$^{+0.02}_{-0.03}$ & 135$^{+20}_{-11}$ & 1.26$^{+0.14}_{-0.13}$ & 1.70$^{+0.11}_{-0.18}$ & 2.89$^{+0.58}_{-0.36}$ & 2.42$^{+0.17}_{-0.12}$ & - & 2.42$^{+0.03}_{-0.07}$ & 1416/1371 \\
& C & - & 1.91$^{+0.11}_{-0.10}$ & 114$^{+114}_{-30}$ & 1.55$^{+0.49}_{-0.50}$ & $<$2.56 & 3.56$^{+2.79}_{-1.47}$ & 2.52$^{+0.45}_{-0.34}$ & - & 1.39$^{+0.14}_{-0.10}$ & 705/740 \\
NGC 5506 & A & 4.07$^{+0.87}_{-0.86}$ & 1.83$^{+0.09}_{-0.07}$ & $>$143 & 0.30$^{+0.20}_{-0.16}$ & 3.00$^{+0.37}_{-0.21}$ & 1.28$^{+1.02}_{-0.39}$ & 2.02$^{+0.64}_{-0.65}$ & 2.13$^{+1.03}_{-1.03}$ & 3.55$^{+0.45}_{-0.82}$ & 1240/1228 \\
& B & 0.72$^{+1.57}_{-0.68}$ & 1.50$^{+0.19}_{-0.06}$ & 81$^{+78}_{-10}$ & 0.66$^{+0.10}_{-0.09}$ & 2.75$^{+0.34}_{-0.29}$ & 5.00$^{+2.02}_{-2.07}$ & 3.31$^{+0.21}_{-0.23}$ & 0.68$^{+0.60}_{-0.14}$ & 2.42$^{+0.16}_{-0.12}$ & 1263/1310 \\
& C & $<$1.94 & 1.54$^{+0.14}_{-0.11}$ & 90$^{+62}_{-21}$ & 0.64$^{+0.11}_{-0.08}$ & 3.00$^{+0.10}_{-0.14}$ & 5.00$^{+4.38}_{-1.55}$ & 3.08$^{+0.25}_{-0.24}$ & 0.56$^{+0.35}_{-0.21}$ & 3.47$^{+0.28}_{-0.25}$ & 1220/1293 \\
MCG+08-11-011 & A & - & 1.83$^{+0.01}_{-0.01}$  & 153$^{+15}_{-13}$  & 0.40$^{+0.06}_{-0.05}$  & - & - & - & 2.91$^{+0.02}_{-0.02}$  & - & 1368/1316 \\
& B & - & 1.76$^{+0.02}_{-0.02}$  & 88$^{+10}_{-9}$  & 0.63$^{+0.15}_{-0.13}$  & - & - & - & 2.18$^{+0.03}_{-0.03}$  & - & 781/745 \\
GRS 1734-292 & A & 1.43$^{+0.45}_{-0.43}$ & 1.66$^{+0.02}_{-0.02}$ & 60$^{+5}_{-4}$ & 0.27$^{+0.08}_{-0.09}$ & - & - & - & 3.46$^{+0.08}_{-0.04}$ & - & 948/865 \\
& B & 2.97$^{+0.43}_{-0.47}$ & 1.68$^{+0.02}_{-0.01}$ & 87$^{+9}_{-5}$ & 0.27$^{+0.11}_{-0.06}$ & - & - & - & 3.00$^{+0.05}_{-0.08}$ & - & 825/856 \\
Mrk 926 & A & - & 1.75$^{+0.01}_{-0.01}$ & 164$^{+17}_{-15}$ & 0.13$^{+0.04}_{-0.04}$ & - & - & - &  3.37$^{+0.02}_{-0.03}$ & - & 1519/1468 \\ 
& B & - & 1.70$^{+0.03}_{-0.03}$ & 136$^{+74}_{-36}$ & $<$0.44 & - & - & - &  1.18$^{+0.14}_{-0.04}$ & - & 475/494 \\
Mrk 841 & A & - & 1.80$^{+0.03}_{-0.03}$ & 134$^{+60}_{-34}$ & 0.40$^{+0.23}_{-0.19}$ & - & - & - &  0.93$^{+0.02}_{-0.02}$ & - & 477/507 \\
& B & - & 1.81$^{+0.02}_{-0.02}$ & 120$^{+25}_{-20}$ & 0.56$^{+0.16}_{-0.14}$ & - & - & - &  0.81$^{+0.01}_{-0.01}$ & - & 736/744 \\
NGC 5273 & A & - & 1.50$^{+0.06}_{-0.05}$ & 71$^{+29}_{-17}$ & 1.06$^{+0.35}_{-0.29}$ & $2.00^{*}$ & $>$7.31 & 2.78$^{+0.26}_{-0.28}$ & - & 0.86$^{+0.09}_{-0.13}$  & 594/574 \\ 
& B & - & 1.23$^{+0.07}_{-0.05}$ & 43$^{+14}_{-8}$ & 0.73$^{+0.34}_{-0.21}$ & $2.00^{*}$ & $>$5.35 & 2.46$^{+0.40}_{-0.73}$ & - & 0.80$^{+0.09}_{-0.08}$  & 466/506 \\
NGC 0985 & A & 7.65$^{+1.56}_{-1.53}$ & 1.90$^{+0.05}_{-0.04}$ & $>$139 & 0.60$^{+0.38}_{-0.31}$ & - & - & - &  0.70$^{+0.03}_{-0.03}$ & - & 303/267 \\
& B & - & 1.85$^{+0.03}_{-0.03}$ & 184$^{+150}_{-58}$ & 0.54$^{+0.27}_{-0.23}$ & - & - & - &  0.86$^{+0.02}_{-0.02}$ & - & 390/466 \\
HE 1143$-$1810 & A & 1.00$^{+0.77}_{-0.77}$ & 1.73$^{+0.03}_{-0.03}$ & 77$^{+18}_{-12}$ & 0.25$^{+0.17}_{-0.15}$ & - & - & - &  1.07$^{+0.09}_{-0.03}$ & - & 491/506 \\
& B & 2.54$^{+0.84}_{-0.81}$ & 1.85$^{+0.03}_{-0.04}$ & 141$^{+71}_{-33}$ & 0.42$^{+0.21}_{-0.19}$ & - & - & - &  1.18$^{+0.03}_{-0.02}$ & - & 469/486 \\
& C & - & 1.77$^{+0.02}_{-0.02}$ & 104$^{+24}_{-17}$ & 0.33$^{+0.15}_{-0.14}$ & - & - & - &  1.37$^{+0.02}_{-0.02}$ & - & 530/597 \\
& D & 1.85$^{+0.71}_{-0.68}$ & 1.85$^{+0.02}_{-0.02}$ & 175$^{+91}_{-45}$ & 0.27$^{+0.16}_{-0.14}$ & - & - & - &  1.56$^{+0.03}_{-0.03}$ & - & 590/561 \\
& E & 1.44$^{+0.68}_{-0.67}$ & 1.82$^{+0.03}_{-0.02}$ & 155$^{+81}_{-36}$ & 0.20$^{+0.15}_{-0.12}$ & - & - & - &  1.48$^{+0.03}_{-0.03}$ & - & 590/568 \\
\hline\hline
\end{tabular}
\end{table*}

\clearpage
\section{Notes on individual sources}
\label{sec:sources}
We discuss here the results of our spectral analysis as obtained from the physically motivated models, const $\times$ TBabs $\times$ zTBabs $\times$ (xillver/relxill/relxill+xillver) (see Table \ref{table-4}) and const $\times$ TBabs $\times$ zTBabs $\times$ (xillverCP/relxillCP/relxillCP+xillverCP) (see Table \ref{table-5}) for $\rm{E_{cut}}$ and $\rm{kT_{e}}$ measurements respectively. We also make a comparison of the results obtained using these two models in this work with those in the literature if available. 

\subsection{1H 0419$-$577}
This bright Seyfert 1 galaxy is known to have a low temperature corona based
on the studies available in literature. The coronal properties of this source
have been investigated by several authors using the observations carried out
by {\it NuSTAR} in 2015 (indicated as epoch C in this work). Using {\it relxill}
 model, \cite{2018MNRAS.476.1258T} found  $\rm{E_{cut}}$ $=$ 63$^{+8}_{-9}$ keV. 
The same authors using {\it COMPTT} and {\it nthcomp} models found 
$\rm{kT_{e}}$ values of  15.1$\pm$0.8 keV and  13.4$\pm$1.0 keV respectively. 
Similarly,  \cite{2021A&A...655A..60A} from modelling of the spectra 
using {\it pexmon} found a value of $\rm{E_{cut}}$ of 54$^{+4}_{-4}$ keV. 
Also, recently, \cite{Kang_2022} from modelling of the spectrum with 
{\it pexrav} found a value of $\rm{E_{cut}}$ $=$ 54$^{+8}_{-6}$ keV. 
Similarly, from {\it relxillCP} model fits to the data, \cite{Kang_2022}
found a value of $\rm{kT_{e}}$ $=$ 16.0$\pm$1.0. 

In this work we analysed three epochs of data, of which the results for epochs 
A and B are reported for the first time.  For epoch C we found $\rm{E_{cut}}$ 
$=$  51$^{+9}_{-4}$ keV and 47$^{+2}_{-2}$ keV from {\it pexrav} and {\it xillver} 
fit. From Comptonization model fit to the observations,  we found $\rm{kT_{e}}$ 
$=$ 15.0$\pm$1.0.  Thus our results for epoch C are in agreement with that of 
\cite{2018MNRAS.476.1258T} and \cite{Kang_2022}. For the other two epochs too,  
we obtained low values for $\rm{E_{cut}}$ and $\rm{kT_{e}}$. The values of 
$\rm{kT_{e}}$ obtained for the three epochs are in agreement within errors, 
and thus from the analysis of the present data sets, we did not find variation 
in $\rm{kT_{e}}$ .

\subsection{Mrk 915}
Mrk 915 was observed three times in December 2014. Of these three epochs, the data
from epoch A has been analysed earlier by \cite{2021MNRAS.506.4960H} and \cite{Kang_2022}.
By jointly fitting {\it NuSTAR} and  {\it XMM-Newton} observations, \cite{2021MNRAS.506.4960H} 
reported a $\rm{E_{cut}}$ of 57.9$^{+11.2}_{-7.4}$ keV. From analysis of {\it NuSTAR} data, 
\cite{Kang_2022} obtained a value of $\rm{kT_{e}}$ > 69 keV. In our analysis too, we
obtained a lower limit of 77 keV for $\rm{kT_{e}}$. For epochs B and C, we obtained values of 
$\rm{kT_{e}}$ = 30$^{+124}_{-12}$ keV  and 17$^{+7}_{-3}$ keV respectively. 
Of the three epochs of data, analysis of data from epochs B and
C are reported for the first time. Due to large error bars, variation of $\rm{kT_{e}}$ if any
could not be detected in this source.

\subsection{3C 111}
This source has been observed by {\it NuSTAR} for three epochs, twice in the
year 2017 (epochs A and B in this work)  and once in the year 2019 (epoch C in this work).    
The data from epochs A and B were analysed earlier by \cite{2020ApJ...901..111K}. 
who found a lower limit of $\rm{E_{cut}}$ $>$ 228 keV in epoch A and 
$\rm{E_{cut}}$ $=$  165$^{+202}_{-47}$ keV in epoch B. Using {\it INTEGRAL} data 
\cite{2014ApJ...782L..25M} obtained a value of  
$\rm{E_{cut}}$ $=$  136$^{+47}_{-29}$ keV for this source. On analysis of
the observations obtained in epoch B, \cite{Kang_2022} obtained values of 
 $\rm{E_{cut}}$ $=$  174$^{+166}_{-57}$ keV and $\rm{kT_{e}}$ $>$  35 keV respectively. 
Recently, \cite{2021A&A...655A..60A} performed a spectral fit to the
observations in epoch C and obtained $\rm{E_{cut}}$ $=$ 148$^{+102}_{-43}$ keV.
Our results on this source are in agreement with the values available in the literature.
Of the three epochs, we obtained lower limits to $\rm{kT_{e}}$ of 63 and 50 keV for epochs
A and C, while for epoch B we obtained a $\rm{kT_{e}}$  of 40$^{+47}_{-12}$ keV. We thus
conclude that we could not find variation in $\rm{kT_{e}}$ in 3C 111.

\subsection{NGC 3783}
This Seyfert type AGN  is known for the variable line of sight column density 
\citep{2019A&A...621A..99M}. It has been observed by {\it NuSTAR} four times 
between August and December, 2016 and we analysed all the observations. Using {\it INTEGRAL} data \cite{2014ApJ...782L..25M} found a value of $\rm{E_{cut}}$ $=$  98$^{+79}_{-34}$ keV. On analysis of epoch A data, \cite{Kang_2022} found $\rm{kT_{e}}$ $>$ 150 keV. In this work, we found lower limits to $\rm{kT_{e}}$ during epochs B, C and D, while for epoch A we obtained $\rm{KT_{e}}$ $=$ 65$^{+107}_{-24}$ keV. As $\rm{KT_{e}}$ could not be constrained for three epochs, variation of  $\rm{kT_{e}}$  if any could not be ascertained in this source. 

\subsection{NGC 7469}
We have seven epochs of data from {\it NuSTAR} observed between June and December, 2015. Of these, using the data for epoch D as reported in this work , \cite{2021A&A...655A..60A} obtained a value of  $\rm{E_{cut}}$ $>$ 244 keV. From our analysis of epoch D data in this work we obtained a lower limt of $\rm{E_{cut}}$ $>$ 463 keV.  From joint fitting of the {\it XMM-Newton} and {\it NuSTAR}  observations in epoch G, \cite{2021MNRAS.506.4960H} obtained a value of $\rm{E_{cut}}$ $=$  112.8$^{+32.8}_{-21.9}$ keV. Using only {\it NuSTAR} data for the same epoch, \cite{Kang_2022} measured a lower limit of $\rm{E_{cut}}$ $>$ 262 keV. For the same epoch the authors found $\rm{kT_{e}}$ $>$ 77 keV. Using {\it BeppoSAX} data \cite{2007A&A...461.1209D} reported a $\rm{E_{cut}}$ $=$  211$^{+235}_{-95}$ keV. Though the source has been observed seven times by {\it NuSTAR}, results on only two epochs of observations (epochs D and G) are available in literature. Our analysis of all the seven epochs of data in a homogeneous manner could yield only lower limits to $\rm{kT_{e}}$, except for epoch A, where we found a value of $\rm{kT_{e}}$ = 43$^{+42}_{-14}$ keV.

\subsection{Mrk 110}
Mrk 110 was observed three times between 2017 and 2020 by {\it NuSTAR}. \cite{2020MNRAS.495.3373E} analysed the epoch A {\it NuSTAR} spectrum to model the reflection features of the spectrum. The authors reported $\rm{E_{cut}}$ $=$ 219.30$^{+437.39}_{-112.36}$ keV and $\rm{E_{cut}}$ $=$ 219.30$^{+412.27}_{-113.95}$ keV for the choice of black hole spin $=$ 0 and 0.998 respectively. From an analysis of epoch A spectrum \cite{2021A&A...655A..60A} estimated $\rm{E_{cut}}$ $=$ 93$^{+13}_{-10}$ keV, while \cite{Kang_2022} reported $\rm{E_{cut}}$ $=$ 160$^{+35}_{-24}$ keV by fitting the same {\it NuSTAR} spectrum. The authors also found $\rm{kT_{e}}$ $=$ 57$^{+54}_{-18}$ keV. All the epochs were analysed recently by \cite{2021A&A...654A..89P}. The author did a broad band X-ray spectral analysis using the simultaneous {\it XMM-Newton} and {\it NuSTAR} data. From modelling the source spectra with {\it relxill} for epoch A, B and C respectively the authors reported $\rm{E_{cut}}$ $=$ 117$^{+12}_{-17}$ keV, 113$^{+28}_{-21}$ keV and 126$^{+35}_{-26}$ keV. Using the Comptonization model {\it relxillCP}, the authors obtained $\rm{kT_{e}}$ $=$ 26$^{+4}_{-3}$ keV, 26$^{+8}_{-5}$ keV and 26$^{+8}_{-5}$ keV. Our results on $\rm{kT_{e}}$ for the three epochs are in agreement with that reported by \cite{2021A&A...654A..89P}. 

\subsection{UGC 06728}
UGC 06728 was observed twice by {\it NuSTAR} between July, 2016 and October, 2017. 
From an analysis of the data acquired in the year 2017,\cite{2021A&A...655A..60A} obtained 
a value of   $\rm{E_{cut}}$ $=$ 152$^{+131}_{-50}$ keV.  For the same data set 
\cite{Kang_2022} obtained  $\rm{E_{cut}}$ values of 230$^{+933}_{-108}$ keV and 
183$^{+452}_{-62}$ keV respectively, from fitting two different models to the observations. 
The same authors using {\it relxillCP} model fit to the data obtained a lower limit of 
$\rm{kT_{e}}$ $>$ 26 keV. Our analysis too yeilded a value of  $\rm{kT_{e}}$ $>$ 23 keV 
for epoch B. For epoch A, we obtained a value of $\rm{kT_{e}}$ = 17$^{+34}_{-5}$keV. 
Of the two epochs of data availabe on this source, we could constrain $\rm{kT_{e}}$ 
only for epoch A, and therfore, the variability nature of the temperature of the 
corona if any could not be established. 

\subsection{NGC 4258}
We have {\it NuSTAR} observations  for this  source on two epochs observed 
between November 2015 and January 2016. Results on the analysis of these data 
sets are reported for the first time. Using {\it XMM-Newton} data 
\cite{Vasudevan_2013} has reported a lower limit of $\rm{E_{cut}}$ $>$ 282 keV 
for this source. We obtained values of $\rm{E_{cut}}$ $=$ 41$^{+5}_{-6}$ keV 
and 30$^{+3}_{-3}$ keV respectively for epochs A and B. There is thus
an indication of $\rm{E_{cut}}$ variation in this source. However, from 
{\it xillverCP} model fit to both the epochs of observations we found $\rm{kT_{e}}$ of 13$^{+7}_{-3}$ keV and  9$^{+2}_{-1}$ keV respectively.
Thus, considering the errors, we conclude the temperature of the corona of the 
source to be non variable between November 2015 and January 2016.

\subsection{KUG 1141+371}
This source was observed twice by {\it NuSTAR}, once in December 2019 (Epoch A)  
and again in May 2020 (Epoch B). From model fits to the data, we found lower
limits of $\rm{kT_{e}}$ $>$ 22 keV and  $>$ 16 keV for epochs A and B
respectively. Using {\it XMM-Newton} data \cite{Vasudevan_2013} reported a $\rm{E_{cut}}$ of  263$^{+\infty}_{-212}$ keV. From an analysis of the epoch A data, \cite{2021A&A...655A..60A} obtained  $\rm{E_{cut}}$ $>$ 79  keV. From our analysis, while we found a value of  $\rm{E_{cut}}$ $=$ 133$^{+64}_{-34}$ keV for epoch A, we found a lower limit for epoch B. 

\subsection{MCG-06-30-15}
This source has been studied extensively using {\it INTEGRAL}, {\it BeppoSAX}, 
{\it Swift-XRT}, {\it XMM-Newton} and {\it NuSTAR} data to determine 
$\rm{E_{cut}}$. \cite{2007A&A...461.1209D} reported a $\rm{E_{cut}}$ of 190$^{+110}_{-66}$ keV from an analysis of {\it BeppoSAX} data. Using {\it INTEGRAL} spectra \cite{2014ApJ...782L..25M} reported  $\rm{E_{cut}}$ $=$ 63$^{+24}_{-15}$ keV for the source. \cite{Marinucci_2014} reported a $\rm{E_{cut}}$ $>$ 110 from a joint fit of {\it XMM-Newton} and {\it NuSTAR} data. This source  was observed by {\it NUSTAR} three times on  29 January 2013 (epoch A), 30 January 2013 (epoch B)  and 02 February 2013 (epoch C). Of these three epochs, results are available in literature on epoch B. \cite{2020MNRAS.495.3373E} from fitting of the  epoch B spectrum 
with a combination of {\it relxill+xillver} model found $\rm{E_{cut}}$ $=$ 160.20$^{+28.76}_{-18.26}$ keV for $a_{*}$ $=$ 0.998 and $\rm{E_{cut}}$ $=$ 149.73$^{+31.15}_{-12.46}$ keV for $a_{*}$ $=$ 0.0. From the fitting of same epoch B {\it NuSTAR} data using {\it pexrav} and {\it relxill} \cite{Kang_2022} reported $\rm{E_{cut}}$ $>$ 707 keV and $\rm{E_{cut}}$ $>$ 720 keV respectively. The authors also estimated a high lower limit for $\rm{kT_{e}}$ $>$ 280 keV from the fit of {\it relxillCP} model.  Our analysis of the epoch B {\it NuSTAR} spectrum yielded a $\rm{E_{cut}}$ of 135$^{+20}_{-11}$ keV. 
We obtained $\rm{kT_{e}}$ $=$ 28$^{+59}_{-9}$ keV  for epoch C, while
we obtained lower limits of 50 and 74 keV for epochs A and B respectively. Thus, from our analysis of three epochs of data, we did not find unambiguous evidence of variation in the temperature of the corona in MCG-06-30-15.

\subsection{NGC 5506}
{\it NuSTAR} has observed this source three times on  April 2014 (epoch A), 
December 2019 (epoch B)  and February 2020 (epoch C). From joint fit of the 
epoch A with {\it Swift XRT},  \cite{2015MNRAS.447.3029M} reported a large 
$\rm{E_{cut}}$ $=$ 720$^{+130}_{-190}$ keV.  From a reanalysis of the
same data set,  \cite{2018A&A...614A..37T} found  $\rm{kT_{e}}$ 
$=$ 400 $\pm$ 200 keV. From a joint fit of {\it XMM-Newton} and {\it BAT} data 
\cite{Vasudevan_2013} obtained a relatively lower value of $\rm{E_{cut}}$ $=$ 
166$^{+107}_{-30}$ keV.  \cite{2020ApJ...905...41B} reported rather lower value 
of $\rm{E_{cut}}$ $=$ 110$\pm$10 keV.  Thus, there are discrepant results
available in the literature on this source. From our analysis, we could
constrain both $\rm{E_{cut}}$ and $\rm{kT_{e}}$ on all the three epochs, however, we found no variation in the temperature of the corona.

\subsection{MCG+08-11-011}
We have two epochs of observations  between August 2016 and December 2021. Using 
the {\it INTEGRAL} data \cite{2014ApJ...782L..25M} reported a  $\rm{E_{cut}}$ of 
171$^{+44}_{-30}$ keV. Of the two epochs of  {\it NuSTAR} observations, results
based on the observations carried  out in August 2016 (epoch A) are available in 
literature, while, analysis of observations acquired in December 2021 (epoch B)
has been carried out for the first time.  \cite{2019MNRAS.484.2735M} from joint
spectral fit of  {\it Swift XRT} and epoch A data from {\it NuSTAR} 
obtained a value of $\rm{E_{cut}}$ $=$ 163$^{+53}_{-32}$ keV. Similarly, from spectral analysis of the  {\it NuSTAR} data of epoch A, \cite{2021A&A...655A..60A} found  $\rm{E_{cut}}$ $=$ 140$^{+29}_{-21}$ keV. Analyzing the same epoch of {\it NuSTAR} observation \cite{2018A&A...614A..37T} reported $\rm{E_{cut}}$ $=$ 175$^{+110}_{-50}$ keV. The authos also found $\rm{E_{cut}}$ $=$ 57$^{+60}_{-30}$ keV from fitting the source spectrum using {\it nthcomp}. From our analysis, we found variation in our derived values of both $\rm{E_{cut}}$ and $\rm{kT_{e}}$ between the two epochs. We thus conclude that we have detected $\rm{kT_{e}}$  variation in this source at the 90\% confidence level.

\subsection{GRS 1734-292}
We have two epochs of observations, namely epoch A in September 2014 and epoch B in May 2018. 
Of these, results based on epoch A data are available in literature. 
From joint analysis of {\it INTEGRAL}, {\it XMM} and {\it BAT} data, 
\cite{2014ApJ...782L..25M} obtained a value of $\rm{E_{cut}}$ $=$ 58$^{+24}_{-7}$ keV. 
From a broad band analysis of {\it XMM-Newton} and {\it NuSTAR} data obtained during
epoch A,  \cite{2017MNRAS.466.4193T} found a value of  $\rm{E_{cut}}$ $=$ 53$^{+11}_{-8}$ keV 
and $\rm{kT_{e}}$ $=$ 11.9$^{+1.2}_{-0.9}$ keV. Similarly, from broad band
spectral analysis of {\it Swift XRT} and epoch A data from {\it NuSTAR}, 
\cite{2019MNRAS.484.2735M} obtained a value of $\rm{E_{cut}}$ $=$ 53$^{+13}_{-9}$ keV. 
Using only {\it NuSTAR} data of epoch A, we obtained a value of
$\rm{E_{cut}}$ $=$ 60$^{+5}_{-4}$ keV, which is similar to that obtained by
\cite{2017MNRAS.466.4193T} and \cite{2019MNRAS.484.2735M}. For epoch B, we 
obtained a value of $\rm{E_{cut}}$ $=$ 87$^{+9}_{-5}$ keV. There is an
indication of variation in $\rm{E_{cut}}$, between epochs A and B. However, for 
epochs A and B, we obtained $\rm{kT_{e}}$ of 17$^{+2}_{-2}$ keV and
20$^{+4}_{-2}$ keV respectively, arguing for no coronal temperature 
variation between the two epochs.

\subsection{Mrk 926}
{\it NuSTAR} observed this source two times on November, 2016 (epoch A) and July, 2021 (epoch B). \cite{2021MNRAS.506.4960H}, from analysis of epoch A {\it NuSTAR} spectrum reported $\rm{E_{cut}}$ $=$ 172.8$^{+36.2}_{-26.4}$ keV, while, \cite{Kang_2022} found $\rm{E_{cut}}$ $=$ 323$^{+241}_{-96}$ keV and $\rm{E_{cut}}$ $=$ 292$^{+178}_{-87}$ keV from {\it pexrav} and {\it relxill} model fit to the epoch A spectrum. The authors also reported a $\rm{kT_{e}}$ $>$ 83 keV. From an analysis of simultaneous {\it XMM-Newton} and {\it NuSTAR} epoch A spectra \cite{2022ApJ...927...42K} reported $\rm{E_{cut}}$ $=$ 432$^{+435}_{-144}$ keV and $\rm{kT_{e}}$ $=$ 10.01$^{+0.15}_{-0.07}$ keV. For epoch A, we obtained a value of $\rm{kT_{e}}$ $=$ 45$^{+39}_{-9}$ keV, while for epoch B, we obtained a lower limit of $\rm{kT_{e}}$ $>$ 20 keV. Results on the analysis of {\it NuSTAR} data for epoch B are reported for the first time.

\subsection{Mrk 841}
Between 2015 and 2022, Mrk 841 was observed two times by {\it NuSTAR}. Of these two epochs, results on the analysis of the epoch A spectrum are available in the literature \citep{2018ApJ...866..124K, 2021A&A...655A..60A, 2021MNRAS.506.4960H, Kang_2022, 2022ApJ...927...42K}, however, results based on the spectrum obtained in epoch B are reported for the first time. For epoch A we found a lower limit to $\rm{kT_{e}}$ of 20 keV, while, for epoch B we obtained $\rm{kT_{e}}$ $=$ 33$^{+22}_{-11}$ keV. \cite{Kang_2022} too reported a lower limit of  $\rm{kT_{e}}$ $>$ 44 keV during epoch A.

\subsection{NGC 5273}
This source was observed by {\it NuSTAR} twice, namely July 2014 (epoch A) and July 2022 (epoch B). Of the two data available on this source, epoch A data have been analysed by various authors, while results on epoch B are reported for the first time. Using {\it NuSTAR} and {\it Swift/XRT} together for epoch A, \cite{2017MNRAS.470.3239P} reported $\rm{E_{cut}}$ of 143$^{+96}_{-40}$ keV and $\rm{kT_{e}}$ $=$ 57$^{+18}_{-11}$ keV. For the same data set, \cite{2021A&A...655A..60A} estimated a value of $\rm{E_{cut}}$ $=$ 115$^{+91}_{-37}$ keV. Recently, from a joint analysis of {\it NuSTAR} and {\it Swift} spectra, \cite{2022ApJ...927...42K} reported a value of $\rm{E_{cut}}$ $>$ 220 keV and $\rm{kT_{e}}$ $=$ 3.58$^{+0.16}_{-0.28}$ keV. We found values of $\rm{kT_{e}}$ $=$ 17$^{+5}_{-3}$ keV and 15$^{+3}_{-3}$ keV for epoch A and B respectively.

\subsection{NGC 0985}
{\it NuSTAR} observed this source two times between 2012 and 2021. The epoch A FPMA/FPMB spectrum was obtained on August, 2013 and the epoch B spectrum was taken on September, 2021. The epoch B spectrum of this source is analysed for the first time in this work. Previously, \cite{2018ApJ...866..124K} reported $\rm{E_{cut}}$ $>$ 121 keV from an analysis of epoch A {\it NuSTAR} spectrum. For epochs A and B, we obtained lower limit of $\rm{kT_{e}}$ of 23 keV and 28 keV respectively.

\subsection{HE 1143$-$1810}
This source was observed by {\it NuSTAR} five times in 2017 jointly with {\it XMM-Newton}. All the observations were analysed by \cite{2020A&A...634A..92U}. From joint spectral fits to {\it NuSTAR} and {\it XMM-Newton data}, \cite{2020A&A...634A..92U} found $\rm{kT_{e}}$ of 13$^{+7}_{-3}$, 13$^{+6}_{-3}$, 25$^{+75}_{-8}$, 20$^{+80}_{-6}$ and 20$^{+70}_{-6}$ keV respectively. From our analysis of only {\it NuSTAR} data we could obtain $\rm{kT_{e}}$ for epochs A (20$^{+13}_{-4}$ keV), epoch C (27$^{+19}_{-7}$ keV) and epoch E (37$^{+269}_{-15}$ keV) respectively, while for epochs B and D we could obtain lower limits of $\rm{kT_{e}}$ of 28 keV. Our results for epochs A, C and E are in agreement with that of \cite{2020A&A...634A..92U}. Our results are consistent with no variation in $\rm{kT_{e}}$ in this source.

\bsp	
\label{lastpage}
\end{document}